\documentclass{article}

\usepackage{arxiv}

\usepackage[utf8]{inputenc} 
\usepackage[T1]{fontenc}    
\usepackage{hyperref}       
\usepackage{url}            
\usepackage{booktabs}       
\usepackage{amsfonts}       
\usepackage{nicefrac}       
\usepackage{microtype}      
\usepackage{lipsum}
\usepackage{graphicx}
\graphicspath{ {./images/} }

\usepackage{amsmath,amssymb,latexsym}
\usepackage{amsthm}
\usepackage{stmaryrd}
\usepackage[usenames,dvipsnames]{color}

\newtheorem{theorem}{Theorem}[section]
\newtheorem{corollary}[theorem]{Corollary}

\newtheorem{lemma}[theorem]{Lemma}
\newtheorem{proposition}[theorem]{Proposition}

\theoremstyle{definition}
\newtheorem{definition}[theorem]{Definition}
\newtheorem{remark}[theorem]{Remark}
\newtheorem{remarks}[theorem]{Remarks}

\newtheorem{example}[theorem]{Example}

\newtheorem*{notation}{Notation}

\usepackage[dvipsnames]{xcolor}
\usepackage{tikz}
\usepackage{enumitem}
\usepackage{lineno}
\newcommand{\FF}{\mathbb{F}}
\newcommand{\LL}{\mathbb{L}}
\newcommand{\KK}{\mathbb{K}}
\newcommand{\cv}{{\bf c}}
\newcommand{\Cc}{\mathcal{C}}
\newcommand{\Dc}{\mathcal{D}}
\newcommand{\defegal}{\overset{{\rm def}}{=}}

\title{On the rank weight hierarchy of $M$-codes}

\author{
 Grégory Berhuy \\
  Université Grenoble Alpes\\
  Institut Fourier\\
  CS 40700, 38058 Grenoble cedex 9 \\
  \texttt{gregory.berhuy@univ-grenoble-alpes.fr} \\
   \And
 Julien Molina \\
  Université Grenoble Alpes\\
  Institut Fourier\\
  CS 40700, 38058 Grenoble cedex 9 \\
  \texttt{julien.molina@univ-grenoble-alpes.fr} \\
}

\begin{document}
\maketitle
\begin{abstract}
We study the rank weight hierarchy of linear codes which are stable under a linear endomorphism defined over the base field, in particular when the endomorphism is cyclic. In this last case, we give a necessary and sufficient condition for such a code to have first rank weight equal to $1$ in terms of its generator polynomial, as well as an explicit formula for its last rank weight. 
\end{abstract}


\tableofcontents

\section{Introduction}\label{Section1}
While linear codes, and in particular polynomial and quasi-cyclic codes, are traditionally studied with respect to the Hamming metric, they have increasingly been considered with respect to the rank metric.
Rank metric for (linear) codes was first introduced and used by Gabidulin (1986) \cite{Gabidulin} and Roth (1991) \cite{ROTH}. This metric consists in the following. Let us pick a code $\Cc$ of length $n$ over $\FF_{q^m}$, that is a subset of $\FF_{q^m}^n$. Let us fix an $\FF_q$-basis $\mathcal{B}$ of $\FF_{q^m}$. For any vector $\cv = (c_1,\ldots,c_n)\in \FF_{q^m}^n$, let  $M_{\mathcal{B}}(\cv)\in {\rm M}_{m\times n}(\FF_q)$ be the matrix whose entries are the coordinates of each $c_i$ with respect to $\mathcal{B}$. The \textit{rank weight} of a vector $\cv\in \FF_{q^m}^n$ is thus the rank of $M_{\mathcal{B}}(\cv)$. We then define the \textit{rank distance} between two vectors $\cv$ and $\cv'$ in $\FF_{q^m}^n$ as the rank of the matrix $M_{\mathcal{B}}(\cv-\cv')$. Finally, for a code $\Cc$, we define the \textit{minimum rank distance} to be the minimum of all distances for $\cv\neq \cv'\in \Cc$. In the case where $\Cc$ is a linear code, the previous definition becomes the minimum taken over all the rank weights of $\cv$, for all $\cv\neq 0 \in \Cc$, using the linearity of $\Cc$. We may then consider the \textit{minimum rank weight} or \textit{rank weight} of a linear code $\Cc$, without confusion.  We refer to the recent survey \cite{FnT} and references therein for the state-of-the-art of results on codes with respect to the rank metric and their applications to network coding and cryptography. 

Considering the Hamming weight of a linear code, there exists a sequence of positive integers called generalized Hamming weights that includes the Hamming weight as the first term. The interested reader may refer to \cite[Chapter $7$, Section $10$]{Huffman_Pless_2003} for a quick introduction to this notion.
Similarly to the case of Hamming weights, there exists a sequence of positive integers $M_i$, for all $1\leq i \leq k$, where $k$ is the dimension of the code, called \textit{generalized rank weights}, whose first term of this sequence is the minimum rank weight.
These generalized weights were defined independently by Oggier and Sboui (2012) in \cite{OggierSboui} and by Kurihara, Matsumoto and Uyematsu (2013) in \cite{KuriMatsuUye}. Later, Jurrius and Pellikaan in \cite{JurriusPellikaan} showed that all existing definitions of generalized rank weights are equivalent. In \cite{BerhuyFaselGarotta}, Fasel, Garotta and the first author generalized all these definitions to any arbitrary finite extension of fields and showed their equivalence under the condition $m\geq n$, where $m$ is the degree of the extension and $n$ is the length of the code. Generalized rank weights are also defined using other algebraic structures. In particular, for definitions in connection to matroid theory, we refer to Shiromoto \cite{shiromoto} and to Ghorpade and Johnsen \cite{GhorpadeJohnsen}.

Existing works aiming at understanding the generalized rank weights of linear codes focus on precise classes of linear codes. Among those works, we find the paper of Ducoat and Oggier \cite{DucOggier1} which characterizes when the minimal rank weight of polynomial codes is $1$. Also, the work of Lim and Oggier \cite{Limoggier} on quasi-cyclic codes gives a tighter bound on generalized rank weights than the generalized Singleton bound (see Proposition \ref{PropPoidsgeneralites}) and describes the minimal rank weight for the special case of $1$-generator quasi-cyclic codes. In line with these works, natural questions of classification and characterization of families of codes given the parameters $n,k,q^m$ with respect to the rank metric emerge, akin to the classification of maximum distance separable codes for the Hamming distance. Typical such questions include:
\begin{itemize}
\item
counting the number of linear codes with a given $r$-th generalized rank weight (see Section \ref{Section2} for the relevant definitions) in a specific code family,
\item 
computing or deriving asymptotic results about the density of such codes,
\item
finding new necessary or sufficient conditions on generalized rank weights of a linear code of a specific class to be equal to some fixed values.
\end{itemize}

In this paper, we study linear codes $\Cc\subset \LL^n$ satisfying $\Cc M^t\subset \Cc$ for some matrix $M\in {\rm M}_n(\KK)$, where $\LL/\KK$ is an arbitrary finite extension. Such codes are called {\it M-codes} (see Definition \ref{def-Mcode}), following \cite{lininv}.When $M$ is a cyclic matrix, an $M$-code will be called an {\it $M$-cyclic} code (see Definition \ref{def-Mcyclic}). The family of $M$-codes includes quasi-cyclic codes, while $M$-cyclic codes are a  generalization of polynomial codes (and in particular of cyclic codes). We will answer the previous questions  for mainly the first and the last generalized rank weight of $M$-cyclic codes. More precisely, our work extends previous results on the minimal rank weight given in \cite{DucOggier1} for polynomial codes. Along the way, we will also generalize the results of \cite{Limoggier} for quasi-cyclic codes to $M$-codes.

To get exact computations for all generalized rank weights and arbitrary families of linear codes is a difficult problem.  We thus focus this work on the comprehension of extremal values of the first rank distance, which are $1$ and the Singleton bound, and on the last rank distance. Furthermore, throughout this paper, we highlight that the classification of  $M$-cyclic codes can be understood in terms of the factorizations of the so-called {\it generator polynomial} of the code and of the minimal polynomial $f$ of $M$. Polynomials are objects that we know well and with which we can work easily, in the process echoing the approach to classify cyclic codes.

This paper is organized as follows. Section \ref{Section2} gives all the definitions which we will need throughout the paper. In Section \ref{sec-Mcodes}, we obtain bounds for the rank weight hierarchy of $M$-codes, generalizing the work of Lim and Oggier \cite{Limoggier} for quasi-cyclic codes. We also give a necessary (but not sufficient) condition on the minimal polynomial of $M$ to ensure the existence of an MRD $M$-code (that is, a code $\Cc$ whose first rank weight is maximal).  Section \ref{sec-Mcyclic} deals with $M$-cyclic codes. First, we obtain a closed-form formula for the proportion of $M$-cyclic codes with first rank distance different from $1$, and we characterize the cases where this proportion reaches its extremal values.  These results are then applied in the particular case of cyclic codes over finite fields. Finally, we give a closed-form formula for the last generalized rank weight for an $M$-cyclic code and its dual.

We end this introduction by fixing some notation which will be used throughout this paper.

\begin{notation}
If $a,b\in\mathbb{Z}$, $\left\llbracket a,b\right\rrbracket$ will denote the set $\{ m\in\mathbb{Z}\mid a\leq m\leq b\}$.

Let  $\mathbb{F}$ be a field. Since we deal with linear codes, all the vectors of $\mathbb{F}^n$ will be denoted as row vectors. 

 In particular, if $A\in {\rm M}_{p\times q}(\FF)$, the {\it nullspace} of $A$ will be the subspace $$ \ker(A) \defegal \{\cv\in\FF^q\mid \cv A^t=0\}.$$

 If $M\in{\rm M}_n(\FF)$, the {\it characteristic polynomial} of $M$ is the polynomial $$\chi_M\overset{{\rm def}}{=}\det(xI_n-M)\in\FF[x],$$ while the {\it minimal polynomial} of $M$ is the unique monic polynomial $\mu_M\in\FF[x]$ of least degree such that $\mu_M(M)=0$.

 If $V$ is a finite-dimensional $\FF$-vector space and $u:V\to V$ is an endomorphism of $V$, the characteristic and minimal polynomials of $u$ are defined as the characteristic and minimal polynomials of an arbitrary matrix representation of $u$.  
\end{notation}

\section{Generalized rank weights}\label{Section2}

Traditionally, linear codes are linear subspaces of $\FF_q^n$,  where $\FF_q$ is the finite field with $q$ elements which represents the alphabet in which codeword coefficients live. The rank metric is then studied, as explained in the introduction, on the finite field extension $\FF_{q^m}/\FF_q$ of degree $m$.  The work of Garotta, Fasel and the first author \cite{BerhuyFaselGarotta} showed that the concepts and definitions associated to the rank metric generalize to an arbitrary finite extension. Therefore, from now on, we fix such a  finite extension $\LL/\KK$. This approach is particularly pertinent given the existence of codes designed with the rank metric in mind in arbitrary characteristic \cite{Augot2013,Augot2017}.

{\bf Convention. }For the rest of the paper, unless specified otherwise, $\LL/\KK$ will denote an arbitrary field extension of finite degree $m\geq 1$, and all codes $\Cc$ will be linear $\LL$-subspaces of $\LL^n$, where $m\geq n$ (see \cite[Theorem $5.3$]{BerhuyFaselGarotta}).

We now define the generalized rank weights of a linear code of $\LL^n$.

\begin{definition}[see \cite{JurriusPellikaan}]\label{DefJurriusPelik}
    Let $\Cc$ be an $\LL$-linear code with parameters $[n,k]$, that is a code of length $n$ and dimension $k$.

    Let us pick a $\KK$-basis  $\mathcal{B}$ of $\LL$. For any vector $\cv = (c_1,\ldots,c_n)\in \LL^n$, let  $M_{\mathcal{B}}(\cv)\in {\rm M}_{m\times n}(\KK)$ be the matrix whose entries are the coordinates of each $c_i$ with respect to $\mathcal{B}$. The {\it rank support} of $\cv$, denoted by Rsupp$(\cv)$, is the $\KK$-linear row space of $M_{\mathcal{B}}(\cv)$.  We define wt$_R(\cv)$ to be the dimension of Rsupp$(\cv)$, that is, the rank of $M_{\mathcal{B}}(\cv)$. This does not depend on the choice of $\mathcal{B}$.

    Let $\mathcal{D}$ be an $\LL$-linear subspace of $\Cc$. Then, Rsupp$(\mathcal{D})$, the {\it rank support} of $\mathcal{D}$, is the $\KK$-linear subspace of $\KK^n$ generated by Rsupp$({\bf d})$, for all ${\bf d}\in \mathcal{D}$. Then, wt$_R(\mathcal{D})$ is defined as the dimension of Rsupp$(\mathcal{D})$.

    Finally, for $r\in\left\llbracket 1,k\right\rrbracket$, the {\it $r$-$th$ generalized rank weight} of the code $\Cc$, denoted by $M_r(\Cc)$, is defined as $$M_r(\Cc) \defegal  \underset{\substack{\mathcal{\mathcal{D}} \subset \Cc \\ \dim (\mathcal{\mathcal{D}}) = r}}{\min} \text{wt}_R(\mathcal{\mathcal{D}}).$$ \vspace{0.25cm}
\end{definition}

\begin{remark}\label{remarkcoord}
    For a codeword $\cv = (c_1,...,c_n)\in \Cc$, wt$_R(\cv)$ is nothing but the dimension of the $\KK$-linear subspace generated by the coordinates of $\cv$. In other words, wt$_R(\cv) = \dim_{\KK} \text{Span}(c_1,...,c_n)$.

    Since $\LL$ is a field, hence an integral domain, it is easy to see that, for all $\lambda\in\LL^\times$, and all $x_1,\ldots,x_k\in\LL$, $\lambda x_1,\ldots,\lambda x_k$ are $\KK$-linearly independent if and only if $x_1,\ldots,x_k$ are.

    This equivalence then yields ${\rm wt}_R(\lambda \cv)={\rm wt}_R(\cv)$, and thus $M_1(\LL \cv)={\rm wt}_R(\cv)$.
\end{remark}

The previous remark yields immediately the following well-known result.

\begin{lemma}\label{lemme1motFq}
Let $\Cc\subset \LL^n$ be a linear code. Then, $M_1(\Cc)=1$ if and only if $\Cc\cap \KK^n\neq \{0\}$. \newline
\end{lemma}

\begin{remark}\label{remarkmaxrank}
    By \cite{BerhuyFaselGarotta}, Proposition $4.7$, we get that $\text{wt}_R(\mathcal{\mathcal{D}}) = \dim(\mathcal{D}^*)$, where $\mathcal{D^*}$ is the Galois closure of $\mathcal{D}$, defined in Section $4$ in \cite{BerhuyFaselGarotta} as the intersection of all $\LL$-linear subspaces of $\LL^n$ extended from $\KK^n$ and containing $\Cc$. In particular, for the case of the extension $\FF_{q^m}/\FF_q$, we have $\mathcal{D}^* = \displaystyle\sum_{i=0}^{m-1}\mathcal{D}^{q^i}$, where the power is taken component-wise on the vectors of $\mathcal{D}$. This definition of $\mathcal{D}^*$ generalizes the definition of Galois closure used by Jurrius and Pellikaan in \cite{JurriusPellikaan}.

    Thus, Definition \ref{DefJurriusPelik} may be rewritten as $$M_r(\Cc) = \underset{\substack{\mathcal{\mathcal{D}} \subset \Cc \\ \dim (\mathcal{\mathcal{D}}) = r}}{\min} \dim(\mathcal{D}^*).$$ 

In \cite{BerhuyFaselGarotta}, it is also proved that the definition of the $r$-th generalized rank weight given by Jurrius and Pelikaan in \cite{JurriusPellikaan} is  equivalent to the definition given by Oggier and Sboui in \cite{OggierSboui} which is $$M_r(\Cc) = \underset{\substack{\mathcal{\mathcal{D}} \subset \Cc \\ \dim (\mathcal{\mathcal{D}}) = r}}{\min} \text{maxwt}_R(\mathcal{D}), $$ where maxwt$_R(\mathcal{D})$ is $\underset{d\in \mathcal{D}}{\max}$ wt$_R({\bf d})$.

In particular, if $\Cc$ is a code with parameters $[n,k]$, we have $M_k(\Cc)=\displaystyle\max_{c\in\Cc}{\rm wt}_R({\bf c})$.
\end{remark}

\vspace{0.25cm}
For an $[n,k]$-linear code $\Cc$, the collection of weights $M_1(\Cc), M_2(\Cc), \ldots , M_k(\Cc)$ is called the \textit{rank weight hierarchy of $\Cc$}. In particular, $M_1(\Cc)$ is called the \textit{minimum rank distance/weight}. 

Some well-known properties of the rank weight hierarchy are summarized in the following proposition (the proofs of these properties in the literature are available only in the context of finite fields, but remain true without change for any finite extension $\LL/\KK$).

\begin{proposition}\label{PropPoidsgeneralites}
Let $\Cc$ be an $\LL$-linear code with parameters $[n,k]$.
    \begin{enumerate}
        \item The rank weight hierarchy is  increasing (\cite[Lemma $9$]{KuriMatsuUye}) : $$1\leq M_1(\Cc) < M_2(\Cc) < \cdots < M_k(\Cc) \leq n.$$ 
        \item For $r\in\left\llbracket 1,k\right\rrbracket$, we have a generalized Singleton bound (\cite[Corollary $15$]{KuriMatsuUye}) : $$M_r(\Cc) \leq n - k +r.$$
        \item Let $\Cc^{\perp} = \{ {\bf c}' \in \LL^n \ ; \ \langle {\bf c}',{\bf c}\rangle =0, \forall {\bf c}\in \Cc \}$ be the dual code of $\Cc$, where $\langle \cdot , \cdot \rangle$ is the standard inner product over $\LL^n$. Then (\cite[Theorem I.3]{JeromeDucoat}) : $$\{ M_r(\Cc) \ ; r\in\left\llbracket 1,k\right\rrbracket\} = \{ 1, \ldots , n \}\setminus \{ n+1-M_r(\Cc^{\perp}) \ ; \ r\in\left\llbracket 1,n-k\right\rrbracket\}.$$ 
    \end{enumerate}
\end{proposition}
\vspace{0.25cm}

\begin{definition}\label{MRDdef}
    Let $\Cc$ be an $\LL$-linear code with parameters $[n,k]$. The code $\Cc$ is said to be $r$-MRD (Maximum Rank Distance) if $M_r(\Cc)=n-k+r$, that is, when the $r$-th rank distance reaches the generalized Singleton bound.
    In particular, for $r=1$, we say that $\Cc$ is an MRD code.
\end{definition}

We finish with a useful lemma.

\begin{lemma}\label{Mkpreserv}
Let $P\in{\rm GL}_n(\KK)$, and let $u:\cv\in\LL^n\mapsto \cv P\in\LL^n$.

Then, for all linear $[n,k]$-codes $\Cc$, and for all $r\in\left\llbracket 1,k\right\rrbracket$, we have $M_r(u(\Cc))=M_r(\Cc)$.
\end{lemma}

\begin{proof}
Let us keep the notation of the lemma.
 Since $u$ is an isomorphism of $\LL$-vector spaces, subspaces of $u(\Cc)$ of dimension $r$ have the form $u(\Dc)$ , where $\Dc$ is a subspace of $\Cc$ of dimension $r$. Hence, to prove the desired equality, it is enough to prove that ${\rm wt}_R(u(\Dc))={\rm wt}_R(\Dc)$ for all subspaces $\Dc$ of $\Cc$ of dimension $r$. 
 
 Let us fix a $\KK$-basis $\mathcal{B}$ of $\LL$. 
If $\Dc$ is a subspace of $\Cc$ of dimension $r$ and ${\bf d}\in \Dc$, easy computations show that $M_{\mathcal{B}}(u({\bf d}))=M_{\mathcal{B}}({\bf d})P$. It follows that  ${\rm Rsupp}(u({\bf d}))={\rm Rsupp}({\bf d})P\overset{{\rm def}}{=}\{ {\bf v}P\mid {\bf v}\in {\rm Rsupp}({\bf d})\}$, and thus ${\rm Rsupp}(u(\Dc))={\rm Rsupp}(\Dc)P$ (with similar notation). Since $P\in{\rm GL}_n(\KK)$, we deduce that ${\rm Rsupp}(u(\Dc))$ and ${\rm Rsupp}(\Dc)$ have same dimension over $\KK$, and the desired conclusion follows.
\end{proof}

\section{$M$-codes and their rank weight hierarchy}\label{sec-Mcodes}

\subsection{Definition of $M$-codes}

\begin{definition}\label{def-Mcode}
Let $M\in{\rm M}_n(\KK)$. Following \cite{lininv}, but with slightly different notational conventions, we say that a linear code $\Cc\subset \LL^n$ is an {\it $M$-code} if it is stable under the endomorphism $\rho_M:\cv\in\LL^n\mapsto \cv M^t\in\LL^n$, that is, for all $\cv\in \Cc$, we have $\cv M^t\in\Cc$.

Beware that in \cite{lininv}, the matrix $M$ may have entries in $\LL$, because the authors are investigating generalized Hamming distances. However, the context of our paper is different since we will investigate the rank weight hierarchy of $M$-codes in the sequel. Therefore, we will restrict ourselves to the case where $M\in{\rm M}_n(\KK)$.
\end{definition}

As already mentioned in \cite{lininv}, the family of $M$-codes encompasses various well-known families of codes. To explain how, let us recall a standard notation.

\begin{notation} If $f=x^d+a_{d-1}x^{d-1}+\ldots+a_0\in\KK[x]$, the {\it companion matrix of $f$} is the matrix $$C_f \defegal \left(\begin{array}{ccccc} 0 & 0 & & 0& -a_0 \cr 1 & 0 & & & -a_1 \cr 0 & 1 & & & -a_2 \cr & & \ddots & & \vdots \cr 0 & 0 &\cdots & 1 & -a_{d-1}\end{array}\right)\in{\rm M}_d(\KK).$$
If $d=1$, we then have  $C_f=(-a_0)$.
\end{notation}

With this notation, we see that :

\begin{enumerate}
    \item if $M=C_f$, an $M$-code is an {\it $f$-polynomial code} (in the literature, we may also find the names \textit{polycyclic code}, \textit{pseudo-cyclic code}, or {\it $f$-cyclic code}, see \cite{PolycyclicCode} or \cite{lininv});

    \item if $M=C_f$, where $f=x^n-1, \ x^n+1$ or $x^n-a$ with $a\in \KK$, an $M$-code is a {\it cyclic}, {\it negacyclic} or {\it $a$-constacyclic} code respectively;

    \item if $\ell\geq 1$ and $M=C_f^\ell$, where $f=x^n-1$, an $M$-code is a {\it quasi-cyclic} code.

    \noindent Note that in this case, one may always assume that $\ell\mid n$, after replacing $\ell$ by ${\rm gcd}(\ell,n)$ if necessary.
\end{enumerate}

We continue this subsection by recollecting some facts on decompositions into cyclic subspaces.

Let $\mathbb{F}$ be any field. Recall that an endomorphism $u:E\to E$ of an $\FF$-vector space $E$ of dimension $n$ is {\it cyclic}
if there exists a vector ${\bf v}\in E$ such that $({\bf v}, u({\bf v}),\ldots ,u^{n-1}({\bf v}))$ is an $\FF$-basis of $E$. Such a vector ${\bf v}$ will be called a {\it cyclic vector} for $u$.

A subspace $V$ of $E$ is {\it $u$-cyclic} if it is stable under the endomorphism $u$ and the induced endomorphism on $V$ is cyclic. In other words, a subspace $V$ of $E$ of dimension $d$ is $u$-cyclic if $u(V)\subset V$ and there exists a vector ${\bf v}\in V$ such that $({\bf v}, u({\bf v}),\ldots ,u^{d-1}({\bf v}))$ is an $\FF$-basis of $V$.

Similarly, a matrix $A\in{\rm M}_n(\mathbb{F})$ is ${\it cyclic}$ if the endomorphism $\cv\in \FF^n\mapsto \cv A^t \in\FF^n$ is cyclic, that is, if there exists a vector ${\bf v}\in\FF^n$ such that the family $({\bf v},{\bf v}A^t,\ldots, {\bf v}(A^t)^{n-1})$ is an $\mathbb{F}$-basis of $\FF^n$. Such a vector ${\bf v}$ will be called a {\it cyclic vector} for $A$.

A subspace $V$ of $\FF^n$ is {\it $A$-cyclic} if it is $\rho_A$-cyclic, where  $\rho_A:\cv\in\FF^n\mapsto \cv A^t\in\FF^n$.

It is well-known that $A$ is cyclic if and only if $\mu_A=\chi_A$, where $\mu_A$ is the minimal polynomial of $A$ and $\chi_A$ is the characteristic polynomial of $A$.

Therefore, if $A$ is cyclic, $\mu_A$ has degree $n$. It follows that the map 
$$ev_{\bf v,\mathbb{F} }:\overline{g}\in \mathbb{F}[x]/(\mu_A)\mapsto {\bf v}g(M)^t\in\mathbb{F}^n $$
is an isomorphism of $\mathbb{F}[x]$-modules. Note that this map is even an isomorphism of $\mathbb{F}[x]/(\mu_A)$-modules. 

In particular, a vector ${\bf c}\in\mathbb{F}^n$ may be written in a unique way as ${\bf c}={\bf v}g(M)^t$ for some polynomial $g\in\mathbb{F}[x]$ of degree $\leq n-1$.

Given a matrix $A\in{\rm M}_n(\mathbb{F})$, there exist monic polynomials $\chi_1,\ldots,\chi_t\in\FF[x]$ satisfying $\chi_1\mid\cdots\mid \chi_t$ such that $A$ is similar to the block-diagonal matrix $$\begin{pmatrix}
 C_{\chi_1} & & \cr  & \ddots & \cr & & C_{\chi_t}   
\end{pmatrix}.$$
The polynomials $\chi_1,\ldots,\chi_t\in\FF[x]$ are called the {\it invariant factors} of $A$. They are uniquely determined by $A$ (see \cite[Theorem 7.16]{roman}, except that the order of the invariant factors is reversed). The block-diagonal matrix above is called the {\it Frobenius normal form} or the {\it rational canonical form} of $A$.

In terms of subspaces of $\FF^n$, the previous result yields the existence of
a decomposition $\FF^n=V_1\oplus\cdots\oplus V_t$ such that :

\begin{enumerate}
    \item for $i\in\left\llbracket1,t\right\rrbracket$, $V_i$ is an $A$-cyclic subspace of $\FF^n$;

    \item if $ \chi_i$ is the minimal polynomial of the endomorphism $\cv_i\in V_i\mapsto \cv_iA^t\in V_i$, then $\chi_1\mid \chi_2\mid \cdots\mid \chi_t$.
\end{enumerate}

It is easy to see that $\mu_A=\chi_t$ and $\chi_A=\chi_1\cdots \chi_t$. In particular, $\mu_A\mid \chi_A\mid \mu_A^n$. Hence $\mu_A$ and $\chi_A$ have the same set of monic irreducible factors.

Finally, let $f_1,\ldots,f_s\in\FF[x]$ be the distinct monic irreducible factors of $\chi_A$, and write  $\chi_i=\displaystyle\prod_{j=1}^sf_j^{n_{ij}}, \ n_{ij}\geq 0$.

By \cite[Theorem 7.15]{roman}, if $Q_1,Q_2\in\FF[x]$ are coprime polynomials, then $C_{Q_1Q_2}$ is similar to $\begin{pmatrix}
    C_{Q_1} & \cr & C_{Q_2}
\end{pmatrix}$. 

It follows from the results above that we also have a decomposition $\FF^n=\displaystyle\bigoplus_{i=1}^t\bigoplus_{j=1}^s V_{ij}$, where $V_{ij}$ is an $A$-cyclic 
subspace such that the restriction of $\rho_A$ on $V_{ij}$ has minimal polynomial $f_i^{n_{ij}}$.

We will exploit the existence of these decompositions to give bounds of the rank weight hierarchy of an $M$-code in the next subsection.

We now define a slight generalization of $f$-polynomial codes, namely $M$-cyclic codes. 
As in the case of cyclic codes, these codes are fully determined by their so-called generator polynomials, as we explain now.

\begin{definition}\label{def-Mcyclic}
An $M$-code $\Cc$, where $M\in {\rm M}_n(\KK)$ is a cyclic matrix, will be called an {\it $M$-cyclic} code.
\end{definition}

\begin{example}
Any companion matrix being a cyclic matrix, $f$-polynomial codes (and, in particular, cyclic codes) are $M$-cyclic codes.
\end{example}

\begin{notation}
  If $\mathbb{F}$ is a field and $d\geq 0$ is an integer,  we will denote by $\FF[x]_{<d}$ the subspace of polynomials of $\FF[x]$ with degree $<d$. 
\end{notation}

The following easy lemma will be crucial for the sequel.
 
\begin{lemma}\label{interKn}
Let $\LL/\KK$ be a field extension, and let $M\in{\rm M}_n(\KK)$ be a cyclic matrix. Let ${\bf v}\in \KK^n$ be a cyclic vector for $M$. 

 Then, ${\bf v}$ is also a cyclic vector for $M$, when $M$ is viewed as a matrix with entries in $\LL$. In other words, for all $\cv\in\LL^n$, there is a unique polynomial $g\in \LL[x]_{<n}$ such that $\cv={\bf v}g(M)^t$.
\end{lemma}

\begin{proof}
Let ${\bf v}\in \KK^n$ be a cyclic vector for $M$. Then  $({\bf v},{\bf v}M^t,\ldots, {\bf v}(M^t)^{n-1})$ is a $\KK$-basis of $\KK^n$. The determinant of this family of vectors is a non-zero element of $\KK$, hence a non-zero element of $\LL$. In other words, $({\bf v}, {\bf v}M^t,\ldots, {\bf v}(M^t)^{n-1})$ is also  an $\LL$-basis of $\LL^n$, as required.
\end{proof}

Let $M\in {\rm M}_n(\KK)$ be a cyclic matrix with minimal polynomial $f$, and let ${\bf v}$ be a cyclic vector for $M$. 
Therefore, we have an isomorphism of $\LL[x]/(f)$-modules $$ev_{{\bf v},\LL}:\LL[x]/(f)\overset{\sim}{\to }\LL^n.$$

\begin{remark}\label{remisocyc}
 When $M=\Cc_f$, one may take ${\bf v}=(1, 0,\ldots,0)$, and the basis   $({\bf v},{\bf v}M^t,\ldots, {\bf v}(M^t)^{n-1})$ is nothing but the canonical basis.

If $g=a_{n-1}x^{n-1}+\cdots+a_1x+a_0\in\LL[x]$, the corresponding isomorphism $\LL[x]/(f)\overset{\sim}{\to} \LL^n$ then sends $\overline{g}$ onto $(a_0,\ldots,a_{n-1})$, making this isomorphism canonical.
\end{remark}

By definition, an $M$-cyclic code is nothing but an $\LL[x]/(f)$-submodule of $\LL^n$, so $M$-cyclic codes are in one-to-one correspondence with submodules of $\LL[x]/(f)$. These submodules are just  ideals of $\LL[x]/(f)$. It is well-known that such an ideal has the form $(\overline{g})$, for a unique monic divisor $g\in\LL[x]$ of $f$.

Note that we have a natural isomorphism of $\LL$-algebras $(\LL[x]/(f))/(\overline{g})\simeq\LL[x]/(g)$. It follows that any element of $(\overline{g})$ may be written as $\overline{g Q}$ for a unique polynomial $Q\in\LL[x]$ of degree $<\deg(g)$. 

In particular, $\dim_\LL(\overline{g})=n-\deg(g)$.

All in all, there is a one-to one correspondence between $M$-cyclic codes and monic divisors of $f$ in $\LL[x]$. More precisely, if $g$ is a monic divisor of $f$ of degree $n-k$, the corresponding code is $$\Cc_g=\{ {\bf v}g(M)^t Q(M)^t\mid Q\in\LL[x]\}=\{ {\bf v}g(M)^tQ(M)^t\mid Q\in\LL[x]_{<k}\}.$$
Moreover, $\dim_\LL(\Cc_g)=k$.

Note that, contrary to what it seems, $\Cc_g$ does not depend on the choice of ${\bf v}$.
Indeed, if ${\bf v}_1$ and ${\bf v}_2$ are two cyclic vectors for $M$, then there exist $Q_1,Q_2\in \LL[x]$ such that ${\bf v}_1={\bf v}_2 Q_1(M)^t$ and ${\bf v}_2={\bf v}_1 Q_2(M)^t$. It readily follows that the sets $\{ {\bf v}_ig(M)^t Q(M)^t\mid Q\in\LL[x]\}$, for $i=1,2$, are equal.

If $\Cc$ is an $M$-cyclic code, the unique corresponding monic divisor $g$ will be called the {\it generator polynomial} of $\Cc$ (note that, when $\Cc$ is a cyclic code in the classical sense, we recover the usual definition of the generator polynomial).

\begin{remark}
If $\chi_1,\ldots,\chi_t\in\KK[x]$ are the invariant factors of $M$, then $\LL^n$ and $R\defegal \displaystyle\prod_{i=1}^t\LL[x]/(\chi_i)$ are isomorphic as $\LL[x]$-modules. Hence, there is a one-to-one correspondence between the set of $M$-codes and the set of submodules of $R$ (see \cite{lininv}). This is already well-known for cyclic codes, and more generally for $f$-polynomial codes, as well as for quasi-cyclic codes. Indeed, in the last case, if $M=C_{x^n-1}^\ell$, where $\ell\mid n$, there are $\ell$ invariant factors,  all equal to $x^{n_0}-1$, where $n=n_0\ell$, and a quasi-cyclic code may be seen as a submodule of $\left(\LL[x]/(x^{n_0}-1)\right)^\ell$.
\end{remark}

The $\LL[x]$-module point of view has been used successfully by Lim and Oggier in \cite{Limoggier} to get bounds on the rank weights of quasi-cyclic codes. 
Using  the Chinese Remainder Theorem back and forth multiple times, they prove that a quasi-cyclic code may be decomposed as a direct sum (in the sense of \cite{MartinezP}) of quasi-cyclic codes of smaller lengths, and apply 
\cite[Section III, Corollary $1$]{MartinezP} to get some results on the rank weight hierarchy of the code.
However, their method makes the precise identification of these subcodes quite tricky.

We now propose to clarify and generalize their results using only linear algebra. This is the goal of the next subsection.

\subsection{An upper bound for the rank weight hierarchy of $M$-codes}~

{\bf Notation. }Let $\LL/\KK$ be an arbitrary field extension. If $V$ is a linear subspace of $\KK^n$, we denote by $V_\LL$ the linear subspace of $\LL^n$ generated by the elements of $V$.

The following lemma summarizes the properties of $V_\LL$ we will need in the sequel.

\begin{lemma}\label{lemext}
Let $\LL/\KK$ be an arbitrary field extension. Then, the following properties hold.

\begin{enumerate}
    \item For any subspace $V$ of $\KK^n$, a $\KK$-basis of $V$ is also an $\LL$-basis of $V_\LL$. In particular, $\dim_\LL(V_\LL)=\dim_\KK(V)$.
\\
    \item If $\varphi:V_1\to V_2$ is a $\KK$-linear map (where $V_i$ is a linear subspace of $\KK^{d_i})$, there exists a unique $\LL$-linear map $\varphi_\LL:(V_1)_\LL\to (V_2)_\LL$ such that $\varphi_\LL(v_1)=\varphi(v_1)$ for all $v_1\in V_1$. 

\smallskip

Moreover, for any $\KK$-bases $\mathcal{B}_1$ and $\mathcal{B}_2$ of $V_1$ and $V_2$ respectively, the corresponding matrix representations of $\varphi$ and $\varphi_\LL$ are equal. 
\\
    \item If $\KK^n=V_1\oplus\cdots\oplus V_t$, then $\LL^n=(V_1)_\LL\oplus\cdots\oplus (V_t)_\LL$.
\end{enumerate}

\end{lemma}

\begin{proof}
Note that, if $A\in {\rm M}_{p\times q}(\KK)$, its rank over $\KK$ equals its rank over $\LL$. Indeed a $k\times k$-minor of $A$ is non-zero in $\KK$ if and only if it is non-zero in $\LL$. It follows that a family of $\KK$-linearly independent vectors of $\KK^n$ is also a family of $\LL$-linearly independent vectors of $\LL^n$.

Now, if $(e_1,\ldots,e_d)$ is a $\KK$-basis of $V$, then it spans $V_\LL$ by definition. But $e_1\ldots,e_d$ are $\LL$-linearly independent by the previous observation, and  item 1. follows.

Let us prove item 2. The uniqueness of $\varphi_\LL$ comes from the fact that the elements of $V_1$ span $(V_1)_\LL$ as an $\LL$-vector space. For the existence of $\varphi_\LL$, pick a $\KK$-basis $\mathcal{B}_1$ of $V_1$, and set $\varphi_\LL(v_1)=\varphi(v_1)$ for all $v_1\in\mathcal{B}_1$. By item 1., $\mathcal{B}_1$ is also an $\LL$-basis of $(V_1)_\LL$, hence the previous equalities completely determine $\varphi_\LL$. The last part is then clear.

Finally, assume that $\KK^n=V_1\oplus\cdots\oplus V_t$. For $i\in\left\llbracket 1,t\right\rrbracket$, let $\mathcal{B}_i$ be a $\KK$-basis of $V_i$. Then, their union $\mathcal{B}$ is a $\KK$-basis of $\KK^n$, hence an $\LL$-basis of $\LL^n$. Since $\mathcal{B}_i$ is also an $\LL$-basis of $(V_i)_\LL$, $\mathcal{B}$ is the union of bases of $(V_1)_\LL,\ldots, (V_t)_\LL$. It follows that $\LL^n=(V_1)_\LL\oplus\cdots\oplus (V_t)_\LL$.
\end{proof}

\begin{example}\label{exker}
 Let $\LL/\KK$ be an arbitrary field extension, and let $\varphi: V_1\to V_2$ be a $\KK$-linear map, where $V_i$ is a linear subspace of $\KK^{d_i}$. Then, $\ker(\varphi_\LL)=\ker(\varphi)_\LL$.
 
 Indeed, let $A\in {\rm M}_{p\times q}(\KK)$ be a fixed matrix representation of $\varphi$. By the previous lemma, this is also the matrix representation of $\varphi_\LL$ with respect to the same bases.  Now, $A$ has same rank when viewed as a matrix with entries in $\KK$ or $\LL$, as already observed in the proof of the previous lemma. It follows that $\dim_\LL(\ker(\varphi_\LL))=\dim_\KK(\ker(\varphi))=\dim_\LL(\ker(\varphi)_\LL)$. The inclusion $\ker(\varphi)_\LL\subset \ker(\varphi_\LL)$ being clear, this yields the desired equality.

 In particular, if $M\in {\rm M}_n(\KK)$ and $Q\in\KK[x]$, we have $\ker(Q(M))=\ker(Q(M))_\LL$, where $Q(M)$ is considered as a matrix of ${\rm M}_n(\LL)$ on the left-hand side, and as a matrix of ${\rm M}_n(\KK)$ on the right-hand side.
\end{example}

We now introduce the settings in which we will work in this subsection.

 {\bf Settings. }Let $M\in {\rm M}_n(\KK)$. Assume that $\KK^n=V_1\oplus\cdots V_t$, where each $V_i$ is a $\KK$-linear subspace which is stable under right multiplication by $M^t$.
It is then clear that each $(V_i)_\LL $ is also stable under right multiplication by $M^t$.

For all $i\in\left\llbracket 1,t\right\rrbracket$, let $d_i=\dim_\KK(V_i)=\dim_\LL((V_i)_\LL)$, and let $P_i\in{\rm M}_{d_i\times n}(\KK)$ be a full-rank matrix whose rows form a $\KK$-basis of $V_i$.

If $u_i: \cv_i\in\KK^{d_i}\mapsto \cv_i P_i\in V_i$, the map $(u_i)_\LL:\LL^{d_i}\to(V_i)_\LL$ is then an isomorphism. By assumption on $V_i$, right multiplication by $M^t$ induces a $\KK$-linear endomorphism $\rho_{M,V_i}$ of $V_i$. 

The map $(u_i)_\LL^{-1}(\rho_{M,V_i})_\LL(u_i)_\LL$ is then an automorphism of $\LL^{d_i}$, which is nothing but $(u_i^{-1} \rho_{M,V_i}u_i)_\LL$. In particular, its matrix representation $M_i$ with respect to the canonical basis of $\LL^{d_i}$ is an element of ${\rm M}_{d_i}(\KK)$.

By definition of $M_i$, we then have $(u_i)_\LL(\cv_i)M^t=(u_i)_\LL(\cv_iM_i^t)$ for all $\cv_i\in\LL^{d_i}$.

If $\Cc\subset \LL^n$ is an $[n,k]$-code, we set $$\Cc_i=(u_i)_\LL^{-1}(\Cc\cap (V_i)_\LL)=\{ \cv_i\in\LL^{d_i}\mid \cv_iP_i\in \Cc\cap(V_i)_\LL\}$$ for all $i\in\left\llbracket 1,t\right\rrbracket$.

\begin{lemma}\label{lemci}
 Keeping the previous notation,  for all $i\in\left\llbracket 1, t\right\rrbracket$, the following properties hold:

 \begin{enumerate}
     \item  $(u_i)_\LL$ induces an isomorphism $\Cc_i\simeq \Cc\cap (V_i)_\LL$ which preserves the rank weight hierarchy;

     \item  $\Cc_i$ is an $M_i$-code with parameters $[d_i,k_i]$, where $k_i=\dim_\LL(\Cc\cap(V_i)_\LL)$.
 \end{enumerate}
\end{lemma}

\begin{proof}
The definition of $\Cc_i$ and  Lemma \ref{Mkpreserv} immediately yield item 1. Let $i\in\left\llbracket 1,t\right\rrbracket$, and let $\cv_i\in \Cc_i$, so that $u_i(\cv_i)\in \Cc\cap (V_i)_\LL$. 
Recall that we have  $(u_i)_\LL(\cv_iM_i^t)=(u_i)_\LL(\cv_i)M^t$.
Since $\Cc$ and $(V_i)_\LL$ are stable under right multiplication by $M^t$, it follows that  $(u_i)_\LL(\cv_i M_i^t)\in \Cc\cap (V_i)_\LL$, meaning that $\cv_i M_i^t\in \Cc_i$, as required.
\end{proof}

We are now ready to state the main theorem of this subsection. 

\begin{theorem}\label{weightMgeneral}
Keeping the previous settings, assume that $\Cc\subset\LL^n$ is an $M$-code with parameters $[n,k]$ satisfying $\Cc=(\Cc\cap (V_1)_\LL)\oplus \cdots\oplus (\Cc\cap (V_t)_\LL)$.

Let $\Lambda=\{i\in \left\llbracket 1,t\right\rrbracket \mid \Cc\cap V_i\neq\{0\}\}$.

Then, the isomorphism $u:(\cv_1,\ldots,\cv_t)\in\LL^{d_1}\times\cdots\times \LL^{d_t}\mapsto \displaystyle\sum_{i=1}^t \cv_i P_i\in \LL^n$ induces an isomorphism $$\displaystyle\prod_{i=1}^t\Cc_i\simeq\Cc$$ which preserves the rank weight hierarchy, and for all $r\in\left\llbracket1,k\right\rrbracket$,  we have

$$\begin{array}{lll} M_r(\Cc) &=& \underset{\substack{\sum_{i\in\Lambda}r_i=r \\ r_i\in\left\llbracket 0,k_i\right\rrbracket}}{\min} \displaystyle\sum_{i\in \Lambda}M_{r_i}(\Cc_i) \cr & = & \underset{\substack{\sum_{i\in\Lambda}r_i=r \\ r_i \in\left\llbracket 0,k_i\right\rrbracket}}{\min} \displaystyle\sum_{i\in\Lambda} M_{r_i}(\Cc\cap (V_i)_\LL),\end{array}$$
where $k_i=\dim_\LL(\Cc\cap (V_i)_\LL)$.

Moreover, for all $i\in\Lambda$, and for all $r_i\in\left\llbracket 1,k_i\right\rrbracket$, we have $$ M_{r_i}(\Cc_i)\leq d_i-k_i+r_i.$$

In particular, for all $r\in\left\llbracket 1,k\right\rrbracket$, we have $$M_r(\Cc) \leq\underset{\substack{\sum_{i\in\Lambda}r_i=r \\ r_i \in\left\llbracket 0,k_i\right\rrbracket}}{\min} \displaystyle\big(\sum_{\underset{r_i\neq 0}{i\in \Lambda}}(d_i-k_i)\big) +r.$$

\end{theorem}

\begin{proof}
The isomorphism $u$ is nothing but right multiplication by $P$, where $P\defegal\begin{pmatrix}
\underline{P_1} \cr  \vdots \cr \overline{P_t}
\end{pmatrix}\in{\rm GL}_n(\KK)$. Hence, $u$ preserves the rank weight hierarchy by Lemma \ref{Mkpreserv}.
Now, by construction, we have $$u(\Cc_1\times\cdots\times \Cc_t)=(\Cc\cap (V_1)_\LL)\oplus\cdots\oplus (\Cc\cap (V_t)_\LL)=\Cc,$$
so $\Cc_1\times\cdots\times \Cc_t$ and $\Cc$ are isomorphic and have same weight hierarchy.
Canceling the zero factors preserves the rank weight hierarchy, so $\displaystyle\prod_{i\in\Lambda}\Cc_i$ and $\Cc$ have also same weight hierarchy.
The first equality is then an application of \cite[Section III, Corollary $1$]{MartinezP} to $\displaystyle\prod_{i\in\Lambda}\Cc_i$. The second equality comes from Lemma \ref{lemci}, while the upper bounds are obtained by applying the Singleton bound to $\Cc_i$.
\end{proof}

\begin{corollary}
Keeping the notation of Theorem \ref{weightMgeneral}, we 
have $$M_1(\Cc)=\min_{i\in\Lambda}(M_1(\Cc_i))=\min_{i\in\Lambda}(M_1(\Cc\cap (V_i)_\LL))\leq \underset{i\in\Lambda}{\min} (d_i-k_i)+1, $$

as well as $$M_k(\Cc)=\sum_{i\in\Lambda}M_{k_i}(\Cc_i)=\sum_{i\in\Lambda}M_{k_i}(\Cc\cap (V_i)_\LL)\leq \sum_{i\in\Lambda} d_i. $$
\end{corollary}

\begin{remark}
Theorem \ref{weightMgeneral} shows that an $M$-code $\Cc$ is isomorphic to a code of the form $\Cc_1\times \cdots \times \Cc_t$ with same rank weight hierarchy, where $\Cc_1,\ldots,\Cc_t$ are codes of smaller lengths. 
These codes $\Cc_1,\ldots,\Cc_t$ are the natural generalization of the codes constructed in \cite{Limoggier}. In fact, in the quasi-cyclic case, the former codes and the corresponding upper bounds are exactly the ones obtained in \cite{Limoggier}.

Note that, for all $i\in\left\llbracket 1,t\right\rrbracket$, $\Cc_i$ and $\Cc\cap(V_i)_\LL$ 
are isomorphic and have same weight hierarchy, despite the fact that the latter code has length $n$.

Therefore, in practice, it is enough to compute $\Cc\cap (V_i)_\LL$, or even its dimension $k_i$ if we want to apply the upper bounds provided the theorem and its corollary.
\end{remark}

\begin{corollary}
Keeping the notation of Theorem \ref{weightMgeneral}, let $\Gamma=\{i\in \left\llbracket 1,s\right\rrbracket \mid (V_i)_\LL\subset \Cc\}$, and set $d_\Gamma=\displaystyle\sum_{i\in\Gamma}d_i$.

 Then, for all $r\in\left\llbracket 1, d_\Gamma\right\rrbracket$, we have $M_r(\Cc)=r$.
\end{corollary}

\begin{proof}
 Note that, by definition of $\Gamma$, we have $\Cc\cap (V_i)_\LL=(V_i)_\LL$ and thus $\Cc_i=\LL^{d_i}$ for all $i\in\Gamma$.
 Therefore, $M_{r_i}(\Cc_i)=r_i$ for all $r_i\in\left\llbracket 1,d_i\right\rrbracket$.
 Now, taking $r_i=0$ for all $i\in \Lambda\setminus\Gamma$ in Theorem \ref{weightMgeneral}, we get that $M_r(\Cc)\leq r$ for all $r\in\left\llbracket 1, d_\Gamma\right\rrbracket$. Since the rank weight hierarchy forms an increasing sequence, we get the other inequality.
\end{proof}

\begin{remark}
 In fact, Theorem  \ref{weightMgeneral}  shows that $\Cc$ and $\LL^{d_\Gamma}\times\displaystyle\prod_{i\in\Lambda\setminus \Gamma}\Cc_i$ have same rank weight hierarchy. 

It follows from  \cite[Section III, Corollary $1$]{MartinezP} that $M_r(\Cc)$ is the minimum of the integers $r_1+M_{r-r_1}(\displaystyle\prod_{i\in\Lambda\setminus \Gamma}\Cc_i)$, for all $r\in\left\llbracket 1,k\right\rrbracket$, and all 
 $r_1\in\left\llbracket 0,d_\Gamma\right\rrbracket$. Using the fact that the rank weight hierachy is an increasing sequence, we see that we have $$ M_r(\Cc)=d_\Gamma+M_{r-d_\Gamma}(\displaystyle\prod_{i\in\Lambda\setminus \Gamma}\Cc_i) \ \mbox{ for all }r\in\left\llbracket d_\Gamma, k\right\rrbracket.$$
\end{remark}

We now give two situations for which the previous results may be applied.

{\bf Convention. }In the sequel, if $M\in {\rm M}_n(\KK)$ and $Q\in \KK[x]$, $\ker(Q(M))$ might denote the nullspace of $Q(M)$ in $\KK^n$ or in $\LL^n$. However, the right interpretation will be clear from the context.

\begin{theorem}\label{weightM}
 Let $M\in {\rm M}_n(\KK)$. Let us denote by $\mu_M$ and $\chi_M$ the minimal and characteristic polynomials of $M$ respectively, and write $$\mu_M=f_1^{m_1}\cdots f_s^{m_s} \ \mbox{and } \chi_M=f_1^{n_1}\cdots f_s^{n_s},$$
 where $s, m_i,n_i\geq 1$, and $f_1,\ldots,f_s\in\KK[x]$ are pairwise distinct irreducible monic polynomials.

For $i\in\left\llbracket 1,s\right\rrbracket$, let $d_i\defegal \dim_\KK(\ker(f_i^{m_i}(M)))=n_i\deg(f_i)$, and let $P_i\in{\rm M}_{d_i\times n}(\KK)$ be a full-rank matrix whose rows form a $\KK$-basis of $\ker(f_i^{m_i}(M))$.


Finally, let $\Cc$ be an $M$-code with parameters $[n,k]$, let $k_i\defegal\dim_\LL(\Cc\cap \ker(f_i^{m_i}(M)))$,  and set $$\Cc_i\defegal \{\cv_i\in\LL^{d_i}\mid \cv_iP_i\in\Cc\cap \ker(f_i^{m_i}(M))\}.$$ 

 Then, for all $r\in\left\llbracket 1,k\right\rrbracket$, 
we have 

 
 $$\begin{array}{lll} M_r(\Cc) &=& \underset{\substack{\sum_{i\in\Lambda}r_i=r \\ r_i\in\left\llbracket 0,k_i\right\rrbracket}}{\min} \displaystyle\sum_{i\in \Lambda}M_{r_i}(\Cc_i) \cr & = & \underset{\substack{\sum_{i\in\Lambda}r_i=r \\ r_i \in\left\llbracket 0,k_i\right\rrbracket}}{\min} \displaystyle\sum_{i\in\Lambda} M_{r_i}(\Cc\cap  \ker(f_i^{m_i}(M))),\end{array}$$
 
 where $\Lambda=\{ i\in\left \llbracket 1,s\right\rrbracket \mid  \Cc\cap \ker(f_i^{m_i}(M))\neq \{0\} \}$.

Moreover, for all $i\in\Lambda$, and for all $r_i\in\left\llbracket 1,k_i\right\rrbracket$, we have $$ M_{r_i}(\Cc_i)=M_{r_i}(\Cc\cap \ker(f_i^{m_i}(M)))\leq n_i\deg(f_i)-k_i+r_i.$$

In particular, for all $r\in\left\llbracket 1,k\right\rrbracket$, we have $$M_r(\Cc) \leq\underset{\substack{\sum_{i\in\Lambda}r_i=r \\ r_i \in\left\llbracket 0,k_i\right\rrbracket}}{\min} \displaystyle\left(\sum_{\underset{r_i\neq 0}{i\in \Lambda}}(n_i\deg(f_i)-k_i)\right) +r.$$

\end{theorem}

\begin{proof}
The fact that $d_i=n_i\deg(f_i)$ is a standard result of linear algebra. However, since it seems difficult to find a reference where this result is explicitely mentioned, we sketch a proof for the convenience of the reader. 

Note that we have $\KK^n=\displaystyle\bigoplus_{i=1}^s\ker(f_i^{m_i}(M))=\bigoplus_{i=1}^s\ker(f_i^{n_i}(M)).$ In particular, $$\displaystyle\sum_{i=1}^s\dim_\KK(\ker(f_i^{m_i}(M)))=\displaystyle\sum_{i=1}^s\dim_\KK(\ker(f_i^{n_i}(M)).$$
Since $\ker(f_i^{m_i}(M))\subset \ker(f_i^{n_i}(M))$, we have $$\dim_\KK(\ker(f_i^{m_i}(M)))\leq \dim_\KK(\ker(f_i^{n_i}(M))),$$ and the previous equality then implies that $\dim_\KK(\ker(f_i^{m_i}(M)))=\dim_\KK(\ker(f_i^{n_i}(M)))$ for all $i\in\left\llbracket1,s\right\rrbracket$.

It is then enough to prove that $\dim_\KK(\ker(f_i^{n_i}(M)))=n_i\deg(f_i)$ for all $i\in\left\llbracket1,s\right\rrbracket$.

Assume first that $f_i=x-\lambda_i\in\KK[x]$ for all $i\in\left\llbracket 1,s\right\rrbracket$, so that $\lambda_i$ is an eigenvalue of $M$ with multiplicity $n_i$. In this case, we need to show that $\dim_\KK(\ker(M-\lambda_i I_n)^{n_i})=n_i$ for all $i\in\left\llbracket1,s\right\rrbracket$. By \cite[Theorem 8.7]{roman}, $M$ is similar to an upper triangular matrix. Comparing characteristic polynomials shows that the diagonal entries of this triangular matrix are precisely the eigenvalues of $M$, and that they appear with the right multiplicity. The desired equality then comes from a direct block-matrix computation.

To get the result in full generality, let $\overline{\KK}$ be an algebraic closure of $\KK$. For $i\in\left\llbracket 1,s\right\rrbracket$, write $f_i=\displaystyle\prod_{j=1}^{e_i}(x-\lambda_{ij})^{t_{ij}}$, where  $\lambda_{ij}\in\overline{\KK}$, and $t_{ij}\geq 1$. 
We then have $$\ker(f_i^{n_i}(M))=\bigoplus_{j=1}^{e_i} \ker(M-\lambda_{ij}I_n)^{n_i t_{ij}},$$ where $M$ is viewed as a matrix with entries in $\overline{\KK}$.
Using the previous case, we get $\dim_{\overline{\KK}}(\ker(f_i^{n_i}(M)))=\displaystyle\sum_{j=1}^{e_i}n_it_{ij}=n_i\deg(f_i)$. Now, we may apply Example \ref{exker} to conclude.

Let us prove the rest of the theorem.

Since the polynomials $f_1^{m_1},\ldots,f_s^{m_s}$ are pairwise coprime, we have $$\KK^n= \ker(f_1^{m_1}(M))\oplus\cdots\oplus \  \ker(f_s^{m_s}(M)).$$ Moreover, each $\ker(f_i^{m_i}(M))$ is stable under right multiplication by $M^t$.

Note that Lemma \ref{lemext} and Example \ref{exker} imply that $$\LL^n= \ker(f_1^{m_1}(M))\oplus\cdots\oplus \  \ker(f_s^{m_s}(M)).$$

We now proceed to prove that $\Cc=\Cc\cap  \ker(f_1^{m_1}(M))\oplus\cdots\oplus \Cc\cap  \ker(f_s^{m_s}(M))$.

The inclusion  $\Cc\cap  \ker(f_1^{m_1}(M))\oplus\cdots\oplus \Cc\cap  \ker(f_s^{m_s}(M))\subset \Cc$ is clear.

Now, for $i\in\left\llbracket 1,s\right\rrbracket$, let $Q_i=\displaystyle\prod_{j\neq i}f_j^{m_j}$. The polynomials $Q_1,\ldots,Q_s$ are globally coprime in $\KK[x]$, so one may write $U_1Q_1+\cdots+U_sQ_s=1$ for some $U_1,\ldots,U_s\in\KK[x]$.

For $\cv\in\Cc$, we then have 
$$\cv= \cv(U_1Q_1)(M)^t+\cdots +\cv(U_sQ_s)(M)^t.$$ By definition, $f_i^{m_i}Q_i$ is equal to $\mu_M$, so $\cv(U_iQ_i)(M)^t$ lies in $\ker(f_i^{m_i}(M))$. But it also belongs to $\Cc$ since $\Cc$ is an $M$-code, hence the desired equality.

Now, apply Theorem \ref{weightMgeneral} to conclude.
\end{proof}

\begin{corollary}\label{coroweightM}


Keeping the notation of Theorem \ref{weightM}, we have
$$M_1(\Cc)=\min_{i\in\Lambda}(M_1(\Cc_i))=\min_{i\in\Lambda}(M_1(\Cc\cap \ker(f_i^{m_i}(M)))\leq \underset{i\in\Lambda}{\min} (n_i\deg(f_i)-k_i)+1, $$

as well as $$M_k(\Cc)=\sum_{i\in\Lambda}M_{k_i}(\Cc_i)=\sum_{i\in\Lambda}M_{k_i}(\Cc\cap \ker(f_i^{m_i}(M)))\leq \sum_{i\in\Lambda} n_i\deg(f_i). $$
\end{corollary}

\begin{corollary}
Let $\Gamma=\{i\in \left\llbracket 1,s\right\rrbracket \mid \ker(f_i^{m_i}(M))\subset \Cc\}$, and set $d_\Gamma=\displaystyle\sum_{i\in\Gamma}d_i$.

 Then, for all $r\in\left\llbracket 1, d_\Gamma\right\rrbracket$, we have $M_r(\Cc)=r$.
\end{corollary}

We now exploit the existence of a  decomposition into cyclic subspaces to relate the rank weight hierarchy of a large family of $M$-codes to the rank weight hierarchy of some polynomial codes.

\begin{theorem}\label{weightMFrob}
 Let $M\in {\rm M}_n(\KK)$, and let $\KK^n=V_1\oplus\cdots\oplus V_t$ be a decomposition of $\KK^n$ into cyclic subspaces such that the restriction of right multiplication by $M^t$ to $V_i$ has minimal polynomial $\Theta_i$.   

Let $D\in \LL[x]$, and let $\Cc=\ker(D(M))$. For $i\in\left\llbracket 1,t\right\rrbracket$, let $d_i=\deg(\Theta_i)$ and let $\Cc_i\subset\LL^{d_i}$ be the $\Theta_i$-polynomial code with generator polynomial $\dfrac{\Theta_i}{{\rm gcd}(D,\Theta_i)}$.

Finally, set $\Lambda=\{i\in\left\llbracket 1,t\right\rrbracket\mid {\gcd}(D,\Theta_i)\neq 1\}$.

Then, for all $i\in \left\llbracket 1,t\right\rrbracket$, we have $k_i\defegal\dim_\LL(\Cc_i)=\deg({\rm gcd}(D,\Theta_i))$, and for all $r_i\in\left\llbracket 1,k_i\right\rrbracket$, we have $$ M_{r_i}(\Cc_i)\leq \deg(\Theta_i)-\deg({\rm gcd}(D,\Theta_i))+r_i.$$

Moreover,
$k\defegal\dim_\LL(\Cc)=\displaystyle\sum_{i\in\Lambda} \deg ({\rm gcd}(D,\Theta_i))$, and for all $r\in\left\llbracket 1,k\right\rrbracket$,
we have $$M_r(\Cc)= \underset{\substack{\sum_{i\in\Lambda}r_i=r \\ r_i\in\left\llbracket 0,k_i\right\rrbracket}}{\min} \displaystyle\sum_{i\in \Lambda}M_{r_i}(\Cc_i).$$
In particular,  for all $r\in\left\llbracket 1,k\right\rrbracket$, we have $$M_r(\Cc) \leq\underset{\substack{\sum_{i\in\Lambda}r_i=r \\ r_i \in\left\llbracket 0,k_i\right\rrbracket}}{\min} \displaystyle\left(\sum_{\underset{r_i\neq 0}{i\in \Lambda}}(\deg(\Theta_i)-\deg({\rm gcd}(D,\Theta_i)))\right) +r.$$

\end{theorem}

\begin{proof}
  Let ${\bf v}_i\in V_i$ be a cyclic vector for the restriction of right multiplication by $M^t$ to $V_i$. Then, $\mathcal{B}_i=({\bf v}_i, {\bf v}_iM^t, \ldots, {\bf v}_i (M^t)^{d_i-1})$ is an $\LL$-basis of $(V_i)_\LL$, and the union $\mathcal{B}$ of $\mathcal{B}_1,\ldots,\mathcal{B}_t$ is an $\LL$-basis of $\LL^n$.

Let us determine $\Cc\cap (V_i)_\LL$. Let $\cv_i={\bf v}_iQ_i(M)^t\in (V_i)_\LL$, where $Q_i\in\LL[x]_{<d_i}$.
Then we have $\cv _iD(M)^t=0$ if and only if ${\bf v}_iQ_i(M)^tD(M)^t=0$, if and only if  $\Theta_i\mid Q_iD$.
Let $D_i={\rm gcd}(D,\Theta_i)$, and write $\Theta_i=D_iA_i$ and $D=D_iB_i$. Then, $\Theta_i\mid Q_iD$ if and only if $A_i\mid Q_iB_i$, if and only if $A_i\mid Q_i$, since $A_i$ and $B_i$ are coprime. It follows that $\Cc\cap (V_i)_\LL=\{ {\bf v}_i A_i(M)^tR_i(M)^t\mid R_i\in\LL[x]_{<d_i-\deg(A_i)}\}$, where $A_i=\dfrac{\Theta_i}{{\rm gcd}(D,\Theta_i)}$.
In particular, $k_i\defegal\dim_\LL(\Cc\cap (V_i)_\LL)=d_i-\deg(A_i)=\deg({\rm gcd}(D,\Theta_i))$.

Let $P_i\in{\rm M}_{d_i\times n}(\KK)$ whose $m^{th}$-row is ${\bf v}_i(M^t)^{m-1}$, and let $u_i:\KK^{d_i}\overset{\sim}{\to}V_i$ be the corresponding isomorphism. 
It is not difficult to see the matrix $M_i$ introduced in the settings is nothing but $C_{\Theta_i}$ (the matrix $P_i$ has been chosen exactly for this purpose). In particular, if ${\bf w}=(1,0,\ldots,0)\in \LL^{d_i}$, for all $ R\in\LL[x]$, we have 
$$(u_i)_\LL({\bf w} R(C_{\Theta_i})^t)=(u_i)_\LL({\bf w})R(M)^t={\bf v}_iR(M)^t.$$
It follows easily that $\Cc_i=(u_i)_\LL^{-1}(\Cc\cap (V_i)_\LL)$ is the $\Theta_i$-polynomial code with generator polynomial $A_i$.

We now check that $\Cc=\Cc\cap (V_1)_\LL\oplus\cdots\oplus \Cc\cap (V_t)_\LL$.

  Let $\cv\in \LL^n$, and let us write $\cv=\displaystyle\sum_{i=1}^t{\bf v}_iQ_i(M)^t$, where $Q_i\in\LL[x]_{<d_i}$.
  Then we have $\cv D(M)^t=0$ if and only if $\displaystyle\sum_{i=1}^t{\bf v}_iQ_i(M)^tD(M)^t=0$.
Since $(V_i)_\LL=\{ {\bf v}_iR(M)^t \mid R\in\LL[x]\}$ and $\LL^n=(V_1)_\LL\oplus\cdots\oplus(V_t)_\LL$, we get that $\cv D(M)^t=0$ if and only if 
${\bf v}_iQ_i(M)^tD(M)^t=0$ for all $i\in\left\llbracket 1,t\right\rrbracket$, that is ${\bf v}_iQ_i(M)^t \in\Cc\cap (V_i)_\LL$ for all $i\in\left\llbracket 1,t\right\rrbracket$.

This shows the desired equality. Now, we may apply Theorem \ref{weightMgeneral} to conclude.  
\end{proof}

\begin{corollary}\label{coroweightMFrob}



Keeping the notation of Theorem \ref{weightMFrob},
we have $$M_1(\Cc)=\min_{i\in\Lambda}(M_1(\Cc_i))\leq \underset{i\in\Lambda}{\min} (\deg(\Theta_i)-\deg({\rm gcd}(P,\Theta_i)))+1, $$

as well as $$M_k(\Cc)=\sum_{i\in\Lambda}M_{k_i}(\Cc_i)\leq \sum_{i\in\Lambda} \deg(\Theta_i). $$
\end{corollary}

\begin{corollary}
Keeping the notation of Theorem \ref{weightMFrob}, let $\Gamma=\{i\in \left\llbracket 1,s\right\rrbracket \mid \Theta_i\mid P\}$, and set $d_\Gamma=\displaystyle\sum_{i\in\Gamma}\deg(\Theta_i)$.

 Then, for all $r\in\left\llbracket 1, d_\Gamma\right\rrbracket$, we have $M_r(\Cc)=r$.
\end{corollary}

\begin{remark}\label{remdecomp}
 The previous theorem and its corollaries may be applied to the case where $\Theta_1=\chi_1,\ldots,\Theta_t=\chi_t$, the invariant factors of $M$.   

If $\chi_M=f_1^{n_1}\cdots f_s^{n_s}$, write $\chi_i=\displaystyle\prod_{j=1}^sf_j^{n_{ij}}, \ n_{ij}\geq 0$.
As already pointed out in the previous subsection, we have a decomposition $\LL^n=\displaystyle\bigoplus_{i=1}^t\bigoplus_{j=1}^s V_{ij}$, where $V_{ij}$ is an $M$-cyclic 
subspace such that the restriction of $\rho_M$ on $V_{ij}$ has minimal polynomial $f_i^{n_{ij}}$. 
We may then also apply our results to this decomposition.
\end{remark}

\begin{example}
 If $M=C_{x^n-1}^\ell$, where $\ell\mid n$, then $M$ has $\ell$ invariant factors, all equal to $x^{n_0}-1$, where $n=n_0\ell $. In particular, $\chi_M=(x^{n_0}-1)^\ell$.

 If ${\rm char}(\KK)$ is prime to $n_0$, the polynomial $x^{n_0}-1$ is separable, and $1$ is a single root of $x^{n_0}-1$. It follows that $\chi_f$ is divisible exactly by $(x-1)^\ell$.
 Theorem \ref{weightM} then shows that we have $M_1(\Cc_1)\leq \ell$.
Hence, we recover Corollary 2 of \cite{Limoggier}.
\end{example}

 \begin{example}\label{exkerpm}


 



 Let $\KK=\FF_5, \LL=\FF_{5^{18}}, f_1=x^2-2, f_2=x^2+x+1$, and set 
$$M=\begin{pmatrix}
   C_{f_1} &  & \cr & C_{f_1f_2^2}& \cr & & C_{f_1^2f_2^3}
\end{pmatrix}\in {\rm M}_{18}(\KK), $$
so that $\chi_1=f_1, \ \chi_2=f_1f_2^2, \ \chi_3=\mu_M=f_1^2 f_2^3$, and $\chi_M=\chi_1\chi_2\chi_3=f_1^4f_2^5$.

We have $f_1=(x-\alpha)(x+\alpha)$ and $f_2=(x-j)(x-j^2)$ in $\LL[x]$ for some suitable $\alpha,j\in\LL$. Taking $D=(x-\alpha)(x-j)$ and $\Cc=\ker(D(M))$, using the formula for the dimension of $\Cc$ given in Theorem \ref{weightMFrob}, we get 

$$ \dim_\KK(\ker(f_1^2(M))=8,  \ \dim_\KK(\ker(f_2^5(M))=10,$$
$$\dim_\LL(\Cc)=5, \ \dim_\LL(\Cc\cap\ker(f_1^2(M)))=3, \ \dim_\LL(\Cc\cap\ker(f_2^3(M)))=2.$$

By Theorem \ref{weightM}, We then have $$M_3(\Cc)=\min(M_3(\Cc_1), M_2(\Cc_1)+M_1(\Cc_2),M_1(\Cc_1)+M_2(\Cc_2))\leq \min (8-3,8-3,10-2)+3,$$
that is, $M_3(\Cc)\leq 8$, while the Singleton bound yields $M_3(\Cc)\leq 16$.
 
Similar computations shows that Theorem \ref{weightMFrob} with $\Theta_1=\chi_1,\Theta_2=\chi_2$ and $\Theta_3=\chi_3$ only yields $M_3(\Cc)\leq 15$. 

Now, let us decompose $\LL^{18}$ as the direct sum of five cyclic subspaces $$\LL^{18}=V_1\oplus V_2\oplus V_3\oplus V_4\oplus V_5, $$ where $$\Theta_1= \Theta_2=f_1, \ \Theta_3=\Theta_4=f_2^2, \ \Theta_5=f_2^3,$$
as in Remark \ref{remdecomp}.

Each $\Cc_i$ has then dimension $1$, and is generated by ${\bf w}_i\defegal{\bf w} (\dfrac{\Theta_i}{{\rm gcd}(D,\Theta_i)})(C_{\Theta_i})^t$, where ${\bf w}=(1,0,\ldots,0)\in\LL^{d_i}$. By Remark \ref{remisocyc}, note that ${\bf w}_i$ is just the vector of coefficients of $\dfrac{\Theta_i}{{\rm gcd}(D,\Theta_i)}$.

We then get that $M_1(\Cc_i)=M_1(\LL {\bf w}_i)={\rm wt}_R({\bf w}_i)$, the last equality coming from Remark \ref{remarkcoord}.
Thus, one may compute $M_r(\Cc)$ for all $r\in\left\llbracket 1,5\right\rrbracket$.

Here, we have $${\bf w}_1={\bf w}_2=(\alpha, 1), \ {\bf w}_3={\bf w}_4=(-j^2,1-j^2,1-j^2,1),$$ as well as  $$\ {\bf w}_5=(-j^2,-2j^2+1,-3j^2+2,-2j^2+3,-j^2+2,1) .$$

Hence, each ${\bf w}_i$ has weight $2$, and Theorem \ref{weightMFrob} yields $M_r(\Cc)=2r$ for all  $r\in\left\llbracket 1,5\right\rrbracket$.
\end{example}

The previous example may be easily generalized as follows.

\begin{theorem}\label{genexfrob}
Let $M\in {\rm M}_n(\KK)$, and let $\KK^n=V_1\oplus\cdots\oplus V_t$ be a decomposition of $\KK^n$ into cyclic subspaces such that the restriction of $\rho_M$ to $V_i$ has minimal polynomial $\Theta_i$.   

  Let $D\in\LL[x]$, and let $\Cc=\ker(D(M))$. Let $\Lambda=\{ i\in\left\llbracket 1,t\right\rrbracket \mid {\rm gcd}(D, \Theta_i)\neq 1\}$.

  Assume that ${\rm gcd}(D,\Theta_i)$ has degree $1$ for all $i\in\Lambda$, and let ${\bf w}_i\in\LL^{d_i}$ be the vector of coefficients of $\displaystyle\frac{\Theta_i}{{\rm gcd}(D,\Theta_i)}$, where $d_i\defegal\deg(\Theta_i)$.

  Then, $\Cc$ has dimension $\vert \Lambda\vert$, and  for all $r\in\left\llbracket 1, \vert\Lambda\vert\right\rrbracket$, we have $$M_r(\Cc)=\underset{\substack{J\subset\left\llbracket 1,\vert\Lambda\vert\right\rrbracket \\ \vert J\vert =r}}{\min} \left(\sum_{j\in J}{\rm wt}_R({\bf w}_j)\right).$$
\end{theorem}


\subsection{A necessary condition for the existence of MRD $M$-codes}

We now derive a necessary condition for the existence of an MRD $M$-code.

\begin{theorem}\label{McodeMRD}
 Let $M\in {\rm M}_n(\KK)$. If there exists an MRD $M$-code $\Cc\neq\LL^n$, then $\mu_M=\pi^\ell$, where $\ell\geq 1$ and $\pi\in\KK[x]$ is a monic polynomial which is irreducible over $\KK$.
\end{theorem}

\begin{proof}
Let $\Cc$ be an $M$-code with parameters $[n,k]$.
Assume that $M_1(\Cc)=n-k+1$. Keeping the notation of Corollary \ref{coroweightM}, let $i_0\in\Lambda$ such that 
$$M_1(\Cc)=M_1(\Cc_{i_0}).$$

We then have $M_1(\Cc)\leq n_{i_0}\deg(f_{i_0})-k_{i_0}+1$.
On the other hand, we have $k=\displaystyle\sum_{i=1}^s k_i$ and $n=\displaystyle\sum_{i=1}^s n_i\deg(f_i).$
It follows that for all $i\neq i_0$, we have $k_i=n_i\deg(f_i)$.
Thus, if $s\geq 2$, one of the $\Cc_i$'s equals $\LL^{k_i}$, and thus satisfies $M_1(\Cc_i)=1$. Corollary \ref{coroweightM} then yields $M_1(\Cc)=1$.
Therefore $n=k$, and thus $\Cc=\LL^n$. Consequently, if there is an MRD $M$-code different from $\LL^n$, then $s=1$. In this case, we have $\mu_M=\pi^\ell$, where $\pi=f_1$ and $\ell=m_1$. 
\end{proof}

\begin{remark}
 The previous necessary condition is obviously not sufficient, as the case of $M=I_n$ already shows.   
\end{remark}

\begin{corollary}
Let $n\geq 2$ be an integer prime to ${\rm char}(\KK)$. Then, there is no MRD $M$-code $\Cc$ different from $\LL^n$ in the following situations: 
    \begin{enumerate}[label=(\roman*)]
        \item $M=C_{x^n-1}^\ell$, where $\ell\mid n$;
        \item $M=C_{x^n+1}$, where $n$ is odd integer;
        \item $M=C_{x^n-a}$, where $a$ is a non-zero $p$-th power in $\KK$ for some prime divisor $p$ of $n$.
    \end{enumerate}
\end{corollary}

\begin{proof}
The minimal polynomial of $M$ is $x^{n_0}-1$ in the first case, where $n=n_0\ell$, $x^n+1$ in the second case and $x^n-a$ in the third case. The assumption on $n$ implies that in all cases, $\mu_M$ is separable. In particular, if $\mu_M=\pi^r$ for some monic irreducible polynomial $\pi\in\KK[x]$ , then $r=1$, that is, $\mu_M$ is irreducible. But none of these three polynomials are irreducible, in view of the various assumptions made in each case. The previous theorem then yields the desired result.
\end{proof}

\begin{remarks}~

\begin{enumerate}
    \item The first item of the previous corollary generalizes the fact that no cyclic code different from $\LL^n$ is MRD (see \cite[Proposition $37$]{DucOggier1}).
\\
    \item If $\KK=\mathbb{F}_q$ and $p$ does not divide the multiplicative order $o(a)$ of $a\in\mathbb{F}_q^\times$, then $a$ is a $p$-th power in $\mathbb{F}_q$. Indeed, by assumption, there exists $u,v\in\mathbb{Z}$ such that $u p+v o(a)=1$. We then easily get that $a=(a^u)^p$.
\\
\\
    In particular, item $(iii)$ of the previous corollary may be applied in this case. For example, if $n$ is coprime to $q(q-1)$, then $a$ is a $p$-th power for any prime divisor $p$ of $n$, and there is no MRD constacyclic code for any $a\in \mathbb{F}_q^\times$ in this case (except $\LL^n$).
    \end{enumerate}
\end{remarks}

\section{The case of $M$-cyclic codes}\label{sec-Mcyclic}

\subsection{$M$-cyclic codes and their rank weight hierarchy.}\label{subsec-Mcycl}

The goal of this short subsection is to apply our previous results to the case of $M$-cyclic codes. In this situation, everything may be translated in terms of generator polynomials. 

\begin{theorem}\label{weightMFrobcyclic}
  Let $M\in {\rm M}_n(\KK)$  be a cyclic matrix, with minimal polynomial $f\in\KK[x]$. Let us write $f=f_1^{m_1}\cdots f_s^{m_s},$
 where $f_1,\ldots,f_s\in\KK[x]$ are pairwise distinct irreducible monic polynomials, and $m_1,\ldots,m_s\geq 1$.

Let $g\in\LL[x]$ be a monic divisor of $f$ of degree $n-k$, and write $g=g_1\cdots g_s$, where $g_i$ is a monic divisor of $f_i^{m_i}$.

For $i\in\left\llbracket 1,s\right\rrbracket$, set $d_i= m_i\deg(f_i)$ and $k_i= m_i\deg(f_i)-\deg(g_i)$. 

Let $\Lambda=\{ i\in\left \llbracket 1,s\right\rrbracket \mid  g_i\neq f_i^{m_i}\}$, and for all $i\in\Lambda$, let $\Cc_i\subset\LL^{d_i}$ be the $f_i^{m_i}$-polynomial code with generator polynomial $g_i$.
Then, for all  $r\in\left\llbracket 1,k\right\rrbracket$, we have $$M_r(\Cc_g)=\underset{\substack{\sum_{i\in\Lambda}r_i=r \\ r_i \in\left\llbracket 0,k_i\right\rrbracket}}{\min} \displaystyle\sum_{i\in\Lambda} M_{r_i}(\Cc_i).$$
 
Moreover, for all $i\in\Lambda$, and for all $r_i\in\left\llbracket 1,k_i\right\rrbracket$, we have $$ M_{r_i}(\Cc_i)\leq \deg(g_i)+r_i.$$

In particular, for all $r\in\left\llbracket 1,k\right\rrbracket$, we have $$M_r(\Cc_g) \leq\underset{\substack{\sum_{i\in\Lambda}r_i=r \\ r_i \in\left\llbracket 0,k_i\right\rrbracket}}{\min} \displaystyle\left(\sum_{\underset{r_i\neq 0}{i\in \Lambda}}(\deg(g_i))\right) +r.$$  
\end{theorem}

\begin{proof}
 Let ${\bf v}\in\KK^n$ be a cyclic vector for $M$, so that $\Cc_g=\{{\bf v}g(M)^tQ(M)^t \mid Q\in \LL[x]\}$.
 
Write $f=gh$. Then $\Cc_g=\ker(h(M))$. Indeed, we have $$\dim_\LL(\ker(h(M)))=\deg({\rm gcd}(h,f))=\deg(h)=n-\deg(g)=\dim_\LL(\Cc_g),$$
as well as the inclusion $\Cc_g\subset \ker(h(M))$, since for all $Q\in\LL[x]$, we have $${\bf v} g(M)^t Q(M)^t h(M)^t={\bf v}Q(M)^t f(M)^t=0.$$
Now, we apply Theorem \ref{weightMFrob} to conclude, after noticing that ${\rm gcd}(h, f_i^{m_i})\neq 1$ if and only if $g_i\neq f_i^{m_i}$.
\end{proof}

\begin{corollary}\label{coroweightMcyclic}
 Keeping the notation of Theorem \ref{weightMFrobcyclic}, we have $$M_1(\Cc_g)=\min_{i\in\Lambda}(M_1(\Cc_i))\leq \underset{i\in\Lambda}{\min} (\deg(g_i))+1, $$

as well as $$M_k(\Cc)=\sum_{i\in\Lambda}M_{k_i}(\Cc_i)\leq \sum_{i\in\Lambda} m_i\deg(f_i). $$
\end{corollary}

\begin{corollary}
Keeping the notation of Theorem \ref{weightMFrobcyclic}, let $\Gamma=\{i\in \left\llbracket 1,s\right\rrbracket \mid g_i=1 \}$, and set $d_\Gamma=\displaystyle\sum_{i\in\Gamma}m_i\deg(f_i)$.

 Then, for all $r\in\left\llbracket 1, d_\Gamma\right\rrbracket$, we have $M_r(\Cc)=r$.
\end{corollary}

\begin{example}\label{exmkMcyclic}

Let $\KK=\FF_3, \LL=\FF_{3^{10}}$, and let $M=C_f\in {\rm M}_9(\KK)$, where $$f=(x^2+1)^2(x+1)^3(x-1)^2.$$ Finally, let $g=(x-i)(x-1)^2\in\LL[x]$, where  $i\in\LL$ satisfies $i^2=-1$. Then $\Cc_g$ has dimension $6$.

Moreover, setting $f_1=x^2+1, f_2=x+1, f_3=x-1$, we have $\Lambda=\{1,2\}$  and $M_1(\Cc_g)\leq 2$ and $M_4(\Cc_g)\leq 7$, while the standard Singleton bound gives $M_1(\Cc_g)\leq 6$ and $M_4(\Cc_g)\leq 9$. 

Then, since $\Gamma =\{ 2\}$, we have $M_r(\Cc_g) =r$ for $r\in \llbracket 1,3\rrbracket$ and $4\leq  M_r(\Cc_g) \leq 7$ for $r\in \llbracket 4,6\rrbracket$.
\end{example}

Theorem \ref{genexfrob} also yields the following result, again noticing that $\Cc_g=\ker(h(M))$, where $f=gh$.

\begin{theorem}
Let $M\in {\rm M}_n(\KK)$  be a cyclic matrix, with minimal polynomial $f\in\KK[x]$. Let us write $f=f_1^{m_1}\cdots f_s^{m_s},$
 where $f_1,\ldots,f_s\in\KK[x]$ are pairwise distinct irreducible monic polynomials, and $m_1,\ldots,m_s\geq 1$.

For $i\in\left\llbracket 1,s\right\rrbracket$, let $d_i=m_i\deg(f_i)$.

Let $g\in\LL[x]$ be a monic divisor of $f$ of degree $n-k$, and write $g=g_1\cdots g_s$, where $g_i$ is a monic divisor of $f_i^{m_i}$.

Let $h\in \LL[x]$ such that $f=gh$. Set $\Lambda=\{ i\in\left \llbracket 1,s\right\rrbracket \mid  g_i\neq f_i^{m_i}\}$.

Assume that ${\rm gcd}(h, f_i^{m_i})$ has degree $1$ for all $i\in\Lambda$, and let ${\bf w}_i\in\LL^{d_i}$ be the vector of coefficients of $g_i$.

  Then, $\Cc$ has dimension $\vert \Lambda\vert$, and  for all $r\in\left\llbracket 1, \vert\Lambda\vert\right\rrbracket$, we have $$M_r(\Cc)=\underset{\substack{J\subset\left\llbracket 1,\vert\Lambda\vert\right\rrbracket \\ \vert J\vert =r}}{\min} \left(\sum_{j\in J}{\rm wt}_R({\bf w}_j)\right).$$
\end{theorem}

\subsection{Counting $M$-cyclic codes with first rank weight equal to $1$ }\label{subsec-P}

The goal of this subsection is to give a characterization of $M$-cyclic codes with first rank weight equal to $1$.
For, according to Lemma \ref{lemme1motFq}, we need to understand the intersection of an arbitrary $M$-cyclic code with $\KK^n$.

This is the content of following result.

\begin{theorem}\label{Mcequal1}
Let $M\in{\rm M}_n(\KK)$ be a cyclic matrix with minimal polynomial $f$. Write  $f=f_1^{m_1}\cdots f_s^{m_s}$, where $f_1,\ldots,f_s\in\KK[x]$ are pairwise distinct monic irreducible polynomials of $\KK[x]$, and $m_1,\ldots,m_s\geq 1$.

Let  $g\in\LL[x]$ be a monic divisor of $f$, and write $g=g_1\cdots g_s$, where $g_i\mid f_i^{m_i}$ in $\LL[x]$.

For $i\in\left\llbracket 1,s\right\rrbracket$, set $\ell_i=\left\lbrace\begin{array}{cl}
  0 &\mbox{ if }g_i=1 \cr \min\{\ell\in\left\llbracket 1,m_i\right\rrbracket\mid\ \  g_i\mid f_i^\ell\}& \mbox{ if }g_i\neq 1 \end{array}\right.$

Finally, let  $d=n-\displaystyle\sum_{i=1}^s\ell_i\deg(f_i)$, and let ${\bf v}\in \KK^n$ be a cyclic vector for $M$.

Then the following properties hold: 

\begin{enumerate}
\item we have $ \displaystyle g \cdot\LL[x] \cap \KK[x] = \prod_{i=1}^sf_i^{\ell_i} \cdot \KK[x]$;
\\
\\
\item 
the map $Q\in\KK[x]_{<d}\mapsto \displaystyle {\bf v}(\prod_{i=1}^sf_i^{\ell_i})(M)^tQ(M)^t\in\Cc_g\cap \KK^n$ is an isomorphism of $\KK$-vector spaces.
\\
\\
\noindent In particular, we have $\dim_\KK(\Cc_g\cap\KK^n)=\displaystyle\sum_{i=1}^s(m_i-\ell_i)\deg(f_i)$;
\\
\item we have $M_1(\Cc_g)=1$ if and only if there exists $i\in \left\llbracket 1,s\right\rrbracket$ such that $\ell_i\leq m_i-1$, that is, such that $g_i\mid f_i^{m_i-1}$.
\end{enumerate}
\end{theorem}
\begin{proof}
Since $f_1,\ldots,f_s$ are irreducible and pairwise distinct, $f_1^{m_1},\ldots,f_s^{m_s}$  are pairwise coprime in $\KK[x]$, and therefore in $\LL[x]$, since the gcd of polynomials is invariant under scalar extension. 
    We may then write $g=g_1\cdots g_s$, where $g_i$ is a monic divisor of $f_i^{m_i}$ in $\LL[x]$.
    
    Let us prove item 1. 
    Note that by definition, we have $g_i\mid f_i^{\ell_i}$ for all $i\in\left\llbracket 1,s\right\rrbracket$.
    This implies that $g\mid \displaystyle\prod_{i=1}^sf_i^{\ell_i}$.
    Hence, $\displaystyle\prod_{i=1}^sf_i^{\ell_i}\in g\cdot\LL[x]\cap\KK[x]$.
    Note now that $g\cdot \LL[x]\cap\KK[x]$ is an ideal of $\KK[x]$ which contains $\displaystyle\prod_{i=1}^sf_i^{\ell_i}$. In particular, it is generated by a monic polynomial $D\in\KK[x]$  dividing $\displaystyle\prod_{i=1}^sf_i^{\ell_i}$. Hence, $D=\displaystyle \prod_{i=1}^sf_i^{r_i}$, where $r_i\in\left\llbracket 0,\ell_i\right\rrbracket$.
    Now, $g\mid D$, so for all $i\in\left\llbracket 1,s\right\rrbracket$, we have $g_i\mid D$, and thus $g_i\mid f_i^{r_i}$.
    If $g_i=1$, then $\ell_i=0$ and thus $r_i=0=\ell_i$. If $g_i\neq 1$, the definition of $\ell_i$ implies that $r_i\geq\ell_i$, and thus $r_i=\ell_i$. All in all, we get $D=\displaystyle\prod_{i=1}^sf_i^{\ell_i}$, as required.

We now prove item 2. Note that, for any $Q\in\LL[x]_{<n}$, we have ${\bf v}Q(M)^t\in\KK^n$ if and only if $Q\in \KK[x]_{<n}$.
Indeed, assume that ${\bf v}Q(M)^t\in\KK^n$. Since ${\bf v} $ is a cyclic vector for $M$, there exists $R\in \KK[x]_{<n}$ such that ${\bf v}Q(M)^t={\bf v}R(M)^t$. By Lemma \ref{interKn}, we have $Q=R\in\KK[x]_{<n}$.
Hence, for all $Q\in \LL[x]_{<n}$, we have ${\bf v} Q(M)^t\in\Cc_g\cap \KK^n$ if and only if $Q\in g\cdot\LL[x]\cap \KK[x]_{<n}$.
By item 1., we have $$ g\cdot \LL[x]\cap \KK[x]_{<n}=(g\cdot \LL[x]\cap\KK[x])\cap\KK[x]_{<n}=   \left(\prod_{i=1}^sf_i^{\ell_i} \cdot \KK[x]\right)\cap \KK[x]_{<n}= \prod_{i=1}^sf_i^{\ell_i}\cdot \KK[x]_{<d}.$$

The $\KK$-linear map $Q\in\KK[x]_{<d}\mapsto \displaystyle {\bf v}(\prod_{i\in I}f_i^{\ell_i})(M)^tQ(M)^t\in\Cc_g\cap \KK^n$ is then surjective. It is also injective since ${\bf v}$ is a cyclic vector for $M$.

The formula for the dimension of $\Cc_g\cap\KK^n$ follows immediately, noticing that we have the equality $\deg(f)=n$, since $M$ is a cyclic matrix.

We finally prove item 3. By Lemma \ref{lemme1motFq}, we have $M_1(\Cc_g)=1$ if and only if $\Cc_g\cap\KK^n\neq \{0\}$, that is, if and only if $\dim_\KK(\Cc_g\cap\KK^n)\neq 0.$ By item 2., this means that $\ell_i\leq m_i-1$ for some $i\in\left \llbracket 1,s\right\rrbracket$, which is equivalent to say that $g_i\mid f_i^{m_i-1}$.
\end{proof}

\begin{corollary}\label{allM11}
Let $M\in{\rm M}_n(\KK)$ be a cyclic matrix with minimal polynomial $f$. Write  $f=f_1^{m_1}\cdots f_s^{m_s}$, where $f_1,\ldots,f_s\in\KK[x]$ are pairwise distinct monic irreducible polynomials of $\KK[x]$, and $m_1,\ldots,m_s\geq 1$.

Then every non-zero $M$-cyclic code has first rank weight equal to $1$ if and only if $f_1,\ldots,f_s$ are irreducible in $\LL[x]$.
\end{corollary}

\begin{proof}
Assume that $f_1,\ldots,f_s$ are irreducible in $\LL[x]$, and let $g\in\LL[x]$ be a divisor of $f$. Then $g=f_1^{r_1}\cdots f_s^{r_s}$, where $r_i\in\left\llbracket 0,m_i\right\rrbracket$, so that $g_i=f_i^{r_i}$  for all $i\in\left\llbracket 1,s\right\rrbracket$. 
Hence, if $\Cc_g\neq\{0\}$, that is, if $g\neq f$,  there exists $i\in \left\llbracket 1,s\right\rrbracket$ such that $g_i\mid f_i^{m_i-1}$. By item 3. of the previous theorem, we get $M_1(\Cc_g)=1$.

Conversely, assume that one of the $f_i$'s is reducible in $\LL[x]$, say $f_1$. Let $p_1\in\LL[x]$ be a divisor of $f_1$ different from $1$ and $f_1$, and set $g=p_1^{m_1} f_2^{m_2}\cdots f_s^{m_s}$. By construction, $g\neq f$, and $\Cc_g$ is non-zero. Now, we have $\ell_i=m_i$ for all $i\in\left\llbracket 1,s\right\rrbracket$, so $M_1(\Cc_g)\neq 1$.
\end{proof}

When $f$ is square-free, with the notation of Theorem \ref{Mcequal1}, we have $\ell_i=0$ if ${\rm gcd}(g,f_i)=1$, and $\ell_i=1$ if ${\rm gcd}(g,f_i)\neq 1$. In this situation, the results of Theorem \ref{Mcequal1} translate as follows.

\begin{theorem}\label{Mcequal1sqfree}
Let $M\in{\rm M}_n(\KK)$ be a cyclic matrix with minimal polynomial $f$. Assume that $f$ is square-free, and write  $f=f_1\cdots f_s$, where $f_1,\ldots,f_s\in\KK[x]$ are pairwise distinct monic irreducible polynomials of $\KK[x]$.

Let  $g\in\LL[x]$ be a monic divisor of $f$, and set $I=\{ i\in\left\llbracket 1,s\right\rrbracket\mid {\rm gcd}(g,f_i)\neq 1\}$. 

Finally, let  $d_I=\displaystyle\sum_{i\notin I}\deg(f_i)$, and let ${\bf v}\in \KK^n$ be a cyclic vector for $M$.

Then the following properties hold: 

\begin{enumerate}
\item We have $ \displaystyle g \cdot\LL[x] \cap \KK[x] = \left(\prod_{i\in I}f_i\right) \cdot \KK[x]$;
\\
\\
\item 
the map $Q\in\KK[x]_{<d_I}\mapsto \displaystyle {\bf v}(\prod_{i\in I}f_i)(M)^tQ(M)^t\in\Cc_g\cap \KK^n$ is an isomorphism of $\KK$-vector spaces.
\\
\\
\noindent In particular, we have $\dim_\KK(\Cc_g\cap\KK^n)=\displaystyle\sum_{i\notin I}\deg(f_i)=n-\sum_{i\in I}\deg(f_i)$;
\\
\item we have $M_1(\Cc_g)=1$ if and only if there exists $i\in \left\llbracket 1,s\right\rrbracket$ such that $g$ is coprime to $f_i$ in $\LL[x]$.
\end{enumerate}
\end{theorem}

We would like now to study the proportion of $M$-codes  $\Cc\subset\LL^n$ with minimal rank distance equal to $1$.

To obtain nicer formulas, we will in fact compute here  codes  $\Cc\subset\LL^n$ with minimal rank distance {\bf  different} from $1$ (this means that either $\Cc=\{0\}$ or $\Cc\neq\{0\}$ and $M_1(\Cc)\geq 2$). We will further assume that the irreducible divisors of the minimal polynomial of $M$ are separable in order to make the statements more enlightening. This condition will be automatically fulfilled when $\KK$ is a perfect field, such as a finite field or a field of characteristic $0$.

We then have the following theorem.

\begin{theorem}\label{thmDenombrement}
      Let $M\in{\rm M}_n(\KK)$ be a cyclic matrix with minimal polynomial $f\in\KK[x]$. Assume that all irreducible divisors of $f$ in $\KK[x]$ are separable, and write  $f=f_1^{m_1}\cdots f_s^{m_s}$, where $f_1,\ldots,f_s\in\KK[x]$ are pairwise distinct separable irreducible  monic polynomials of $\KK[x]$, and $m_1,\ldots,m_s\geq 1$.
      
    \noindent If $D\in\LL[x]$, let $\delta_{\LL}(D)$ be the number of irreducible divisors of $D$ in $\LL[x]$.
    
    \noindent Then, the proportion $\mathbb{P}$ of $M$-cyclic codes  $\Cc\subset\LL^n$ with minimal rank distance different from $1$ is $$\mathbb{P} = \displaystyle \prod_{i=1}^s\left( 1- \left(\dfrac{m_i}{m_i+1}\right)^{\delta_{\LL}(f_i)}\right).$$
    
    In particular, we have $$\prod_{i=1}^s\dfrac{1}{m_i+1}\leq \mathbb{P}\leq \prod_{i=1}^s\left( 1- \left(\dfrac{m_i}{m_i+1}\right)^{\deg(f_i)}\right).$$

Moreover :
    \begin{enumerate}
        \item we have $\mathbb{P}=\displaystyle\prod_{i=1}^s\dfrac{1}{m_i+1}$ if and only if $f_1,\ldots,f_s$ are irreducible in $\LL[x]$, if and only if all non-zero $M$-cyclic codes have first rank weight equal to $1$;
        \\
        \\
        \item we have $\mathbb{P}=\displaystyle \prod_{i=1}^s\left( 1- \left(\dfrac{m_i}{m_i+1}\right)^{\deg(f_i)}\right)$  if and only if $f$ totally splits in $\LL[x]$.
    \end{enumerate}
\end{theorem}

\begin{proof}
First, the number $N$ of $M$-codes $\Cc\subset\LL^n$ equals the number of monic divisors of $f$ in $\LL[x]$. Such a divisor $g$ may be written in a unique way as $g=g_1\cdots g_s$, where $g_i$ is a monic divisor of $f_i^{m_i}$ in $\LL[x]$.
Since $f_i$ is separable, it is the product of $\delta_\LL(f_i)$ pairwise distinct monic irreducible polynomials in $\LL[x]$. Thus, the valuation of each irreducible factor of $f_i^{m_i}$ in $\LL[x]$
 is $m_i$, so that there are exactly $(m_i+1)^{\delta_\LL(f_i)}$ divisors of $f_i^{m_i}$ in $\LL[x]$. Hence, we get $$ N=\prod_{i=1}^s (m_i+1)^{\delta_\LL(f_i)}.$$

By item 3 of Theorem \ref{Mcequal1}, an $M$-code $\Cc$ with generator $g$ satisfies $M_1(\Cc)\neq 1$ if and only if $g_i\nmid f_i^{m_i-1}$ for all $i\in\left\llbracket 1,s\right\rrbracket$. Therefore, the number $N'$ of codes satisfying the required property is $$N'=\prod_{i=1}^s \left((m_i+1)^{\delta_\LL(f_i)}- m_i^{\delta_\LL(f_i)}\right).$$
Since $\mathbb{P}=\dfrac{N'}{N}$, we get the required formula.

To prove the rest of the theorem, note first that, for all $D\in \LL[x]$ of degree $\geq 1$, we have 
$1\leq \delta_\LL(D)\leq \deg(D)$. Moreover, we have 
$\delta_\LL(D)=1$ if and only if $D$ is irreducible in $\LL[x]$, and  $\delta_\LL(D)=\deg(D)$ if and only if $D$  totally splits in $\LL[x]$.

That being said, for all $i\in\left\llbracket 1,s\right\rrbracket$, we have  $1\leq \delta_\LL(f_i)\leq \deg(f_i)$, and thus $$1-\dfrac{m_i}{m_i+1}\leq\left( 1- \left(\dfrac{m_i}{m_i+1}\right)^{\delta_{\LL}(f_i)}\right)\leq\left( 1- \left(\dfrac{m_i}{m_i+1}\right)^{\deg(f_i)}\right) \ \mbox{ for all }i\in\left\llbracket 1,s\right\rrbracket.$$
Multiplying everything yields the desired inequality.

Moreover, the lower bound is attained if and only if we have $\dfrac{m_i}{m_i+1}=\left( \dfrac{m_i}{m_i+1}\right)^{\delta_\LL(f_i)}$ for all $i\in\left\llbracket 1,s\right\rrbracket$, that is $  \delta_\LL(f_i)=1$ for all  $i\in\left\llbracket 1,s\right\rrbracket$, which is equivalent to say that $f_i$ is irreducible in $\LL[x]$ for all \ $i\in\left\llbracket 1,s\right\rrbracket$. This is also equivalent to the fact that  all non-zero $M$-cyclic codes have first rank weight equal to $1$ by Corollary \ref{allM11}.

A similar reasoning shows that the upper bound is attained if and only if each $f_i$ splits completely in $\LL[x]$, which is equivalent to say that $f$ splits completely in $\LL[x]$. 
\end{proof}

\begin{example}
Let $\KK=\mathbb{Q}$ and $\LL=\mathbb{Q}(\zeta_{17})$, where $\zeta_{17}=e^{\frac{2i\pi}{17}}$.

Let $f=(x^2-17)^3(x^3-2)^2$. We then have  $$f_1=x^2+17, \ f_2=x^3-2, \ m_1=3 \mbox{ and }m_2=2.$$ One may check that we have $f_1=(x-\sqrt{17})(x+\sqrt{17})\in\LL[x]$, and that $f_2$ stays irreducible in $\LL[x]$. Therefore, $\delta_\LL(f_1)=2$ and $\delta_\LL(f_2)=1$. We then get $$ \mathbb{P}=\left(1-\left(\dfrac{3}{4}\right)^2 \right)\left(1-\left(\dfrac{2}{3}\right)^1 \right)=\dfrac{7}{48}.$$  

\end{example}

We would like now to apply the previous theorem when $\KK$ and $\LL$ are finite fields. For the rest of this subsection, $q$ is a non-trivial prime power, $\KK=\mathbb{F}_q$ and $\LL=\mathbb{F}_{q^m}$.

First, recall the following lemma (cf. \cite[Theorem 3.46]{LidlNiederreiter}).

\begin{lemma}\label{lemfactofinite}
Let $f\in\FF_q[x]$ be a monic irreducible polynomial, and let $m\geq 1$. Then $f$ factors in $\FF_{q^m}[x]$  as the product of $d$ monic irreducible polynomials of same degree $\dfrac{\deg(f)}{d}$, where $d={\rm gcd}(m,\deg(f))$.
\end{lemma}  

The following corollary is then immediate, taking into account that all irreducible polynomials of $\FF_q[x]$ are separable.

\begin{corollary}\label{coroDenombrementfq}

      Let $M\in{\rm M}_n(\FF_q)$ be a cyclic matrix with minimal polynomial $f\in\FF_q[x]$. Write  $f=f_1^{m_1}\cdots f_s^{m_s}$, where $f_1,\ldots,f_s\in\FF_q[x]$ are pairwise distinct irreducible monic polynomials of $\FF_q[x]$, and $m_1,\ldots,m_s\geq 1$.

    \noindent Then, the proportion $\mathbb{P}$ of $M$-cyclic codes  $\Cc\subset\FF_{q^m}^n$ with minimal rank distance different from $1$ is $$\mathbb{P} = \displaystyle \prod_{i=1}^s\left( 1- \left(\dfrac{m_i}{m_i+1}\right)^{{\rm gcd}(m,\deg(f_i))}\right).$$
    
    In particular, we have $$\prod_{i=1}^s\dfrac{1}{m_i+1}\leq \mathbb{P}\leq \prod_{i=1}^s\left( 1- \left(\dfrac{m_i}{m_i+1}\right)^{\deg(f_i)}\right).$$

Moreover :
    \begin{enumerate}
        \item we have $\mathbb{P}=\displaystyle\prod_{i=1}^s\dfrac{1}{m_i+1}$ if and only if any of the following equivalent conditions is satisfied:
\\        
        \begin {enumerate}
        
        \item ${\rm gcd}(m,\deg(f_i))=1$ for all $i\in\left\llbracket 1,s\right\rrbracket$
\\
        \item $f_i$ is irreducible in $\LL[x]$ for all $i\in\left\llbracket 1,s\right\rrbracket$
  \\      
        \item all non-zero $M$-cyclic codes have first rank weight equal to $1$
        \end{enumerate}
\medskip 
        \item we have $\mathbb{P}=\displaystyle \prod_{i=1}^s\left( 1- \left(\dfrac{m_i}{m_i+1}\right)^{\deg(f_i)}\right)$  if and only if  any of the following equivalent conditions is satisfied:
   \\     
        \begin{enumerate}
         \item $\deg(f_i)\mid m$ for all $i\in\left\llbracket 1,s\right\rrbracket$
\\
    \item $f$ totally splits in $\LL[x]$.
    \end{enumerate}
    \end{enumerate}
\end{corollary}

\begin{example}
    Let $\KK=\FF_3, \LL=\FF_{3^{10}}$, so that $m=10$, and let $$f=(x^2+1)^2(x+1)^3(x-1)^2.$$

    We have $f_1=x^2+1$, $f_2=x+1$, $f_3=x-1$, $m_1=2$, $m_2=3$ and $m_3=2$. Therefore, we get $$\mathbb{P}=\left( 1- \left(\dfrac{2}{3}\right)^2\right)\left( 1- \left(\dfrac{3}{4}\right)^1 \right)\left( 1- \left(\dfrac{2}{3}\right)^1 \right)=\dfrac{5}{108}.$$
\end{example}

We conclude this section by applying this corollary to cyclic codes. To do so, we need some results about the factorization of $x^n-1$ over finite fields. We then recall the following facts. 

\begin{definition}(\cite[Definition 2.44]{LidlNiederreiter})\label{defcycloto}
    Let $q$ be a prime power, and let $n$ be a positive integer coprime to $q$. Let us denote by $\zeta\in\overline{\FF}_q$ a primitive $n$-th root of unity. The {\it $n$-th cyclotomic polynomial} over $\FF_q$ is $$\Phi_{n,\FF_q} = \prod_{\substack{s=1 \\ {\rm gcd}(s,n)=1}}^n(x-\zeta^s).$$
    One may show that $\Phi_{n,\FF_q}\in\FF_q[x]$.
\end{definition}

\begin{theorem}(\cite[Theorems 2.45 and 2.47]{LidlNiederreiter})\label{lemmeordrecycloto}
    Let $q$ be a prime power, and let $n$ be a positive integer coprime to $q$. Then, the following properties hold:
    
    \begin{enumerate}
        \item we have  $x^n-1 = \displaystyle \prod_{d\vert n}\Phi_{d,\FF_q}$;
        \\
        \\
        \item the polynomial $\Phi_{n,\FF_q}$ factors into $\dfrac{\varphi(n)}{o_n(q)}$ pairwise distinct monic irreducible polynomials  of the same degree $o_n(q)$ in $\FF_q[x]$, where $\varphi(n)$ is the Euler totient function and $o_n(q)$ is the multiplicative order of $q$ in $(\mathbb{Z}/n\mathbb{Z})^\times$.
    \end{enumerate}
\end{theorem}

We then get the following result.

\begin{corollary} \label{denombCYCLIC}
    Let $n$ be a positive integer, $q$ a prime power integer, and assume that $gcd(q,n)=1$. 

    Let $ t_{d,q} = \dfrac{\varphi(d)}{o_d(q)}$ and $n_{d,q} = \dfrac{o_d(q)}{o_d(q^m)}={\rm gcd}( o_d(q), m)$, where $\varphi$ is the Euler's totient function and $o_d(q)$ is the multiplicative order of $q$ in $(\mathbb{Z}/d\mathbb{Z})^\times$.

    Finally, set $s=\displaystyle\sum_{d\mid n}t_{d,q}$.

    Then, the proportion of cyclic codes in $\FF_{q^m}[x]/(x^n-1)$ of minimal rank distance different from $1$ is $$\mathbb{P} = \displaystyle \prod_{d\vert n} (1-2^{-n_{d,q}})^{t_{d,q}}.$$
In particular, $$\dfrac{1}{2^s}\leq \mathbb{P}\leq \prod_{d\vert n} (1-2^{-o_d(q)})^{t_{d,q}}$$
Moreover :
    \begin{enumerate}
        \item we have $\mathbb{P}=\dfrac{1}{2^s}$ if and only if any of the following equivalent conditions is satisfied:
\\        
        \begin {enumerate}
        
        \item ${\rm gcd}(m,o_n(q))=1$ 
\\
        \item $f_i$ is irreducible in $\FF_{q^m}[x]$ for all $i\in\left\llbracket 1,s\right\rrbracket$
  \\      
        \item all non-zero cyclic codes have first rank weight equal to $1$
        \end{enumerate}

        \medskip
        
        \item we have $\mathbb{P}=\displaystyle  \prod_{d\vert n} (1-2^{-o_d(q)})^{t_{d,q}}$  if and only if any of the following equivalent conditions is satisfied:
   \\     
        \begin{enumerate}
         \item $o_n(q)\mid m$
\\
    \item $f$ totally splits in $\FF_{q^m}[x]$.
    \end{enumerate}
    \end{enumerate}
\end{corollary}

\begin{proof}
    A cyclic code is just a $C_{x^n-1}$-code, so $f = x^n-1$. 

    Since $n$ is coprime to $q$, we know that $f = x^n-1 = \displaystyle\prod_{d\vert n} \Phi_{d,\FF_q}$. Using Theorem \ref{lemmeordrecycloto}, we get that for all $d\mid n$, 
    $\Phi_{d,\FF_q}$ splits into $t_{d,q}=\dfrac{\varphi(d)}{o_d(q)}$ irreducible polynomials in $\FF_q[x]$, each one of degree of $o_d(q)$.

In particular, $x^n-1$ splits into $s=\displaystyle\sum_{d\mid n} t_{d,q}$ pairwise distinct irreducible factors in $\FF_q[x]$. 

Now, by Lemma \ref{lemfactofinite}, each irreducible factor of degree $d$ splits into ${\rm gcd}(m, o_d(q))=n_{d,q}$ irreducible factors in $\FF_{q^m}[x]$. Everything then follows from Corollary \ref{coroDenombrementfq}, after noticing that $o_d(q)\mid o_n(q)$ for all $d\mid n$ in order to get the last part.
\end{proof}

\begin{example}
    Let $q=3$ and $m=n=8$. We have $$ o_1(q)= 1, \ o_2(q)=1, \ o_4(q)= 2 \ \mbox{ and }o_8(q)=2.$$ Therefore, we get  

     $$ t_{1,q}=1 , \ t_{2,q}=1 , \ t_{4,q}=1 \  \mbox{ and } t_{8,q} =2, $$ as  well as 
     $$ n_{1,q}=1 , \ n_{2,q}=1 , \ n_{4,q}=2 \ \mbox{ and } n_{8,q} =2. $$ 
    Consequently, we have $$ \mathbb{P}=(1-2^{-1})^1(1-2^{-1})^1(1-2^{-2})^1(1-2^{-2})^2=\dfrac{27}{256}.$$ 
    
\end{example}

\begin{remark}\label{remarqueConsta}
   It would be tempting to apply the result of this section to other families of codes, such as constacyclic codes. 
   In \cite[Theorem $18$]{Graner}, an explicit factorization of $x^n-a\in\FF_q[x]$ is proposed when $n$ is coprime to $q$. However, the formula is quite cumbersome, and the degrees of the various irreducible factors, as well as the number of irreducible factors of prescribed degree, do not seem very easy to compute. Anyway, the resulting formula for $\mathbb{P}$ would be probably complicated and not very enlightening.

   However, the case of negacyclic codes may be handled quite easily. Indeed, if $n\geq 1$ is an integer such that $2n$ is coprime to $q$, then we have $x^{2n}-1=(x^n-1)(x^n+1)$. If $n=2^rn'$, where $n'$ is odd, it is then easy to deduce that $x^n+1=\displaystyle\prod_{d'\mid n'}\Phi_{2^{r+1}d',\FF_q}$. Reasoning as in the case of cyclic codes, one may obtain results similar to those described in Corollary  \ref{denombCYCLIC}. 
   This is particularly easy when $n$ is odd, since in this case we have $x^n+1=\displaystyle\prod_{d\mid n}\Phi_{2d,\FF_q}=\displaystyle\prod_{d\mid n}\Phi_{d,\FF_q}(-x),$ and the conclusion of Corollary \ref{denombCYCLIC} holds without change.
   Details are left to the reader.
\end{remark}

\subsection{An explicit formula for the last rank distance of $M$-cyclic codes}\label{subsec-Mk}

The goal of this short section is to compute the last generalized rank distance of an $M$-cyclic code.

We start with a lemma, which is valid for arbitrary linear codes.

\begin{lemma}\label{lemmedernierpoids}
    For any linear code $\Cc \subset \LL^n$ with parameters $[n,k]$, we have $$M_k(\Cc) = n - \dim_{\KK}(\Cc^{\perp}\cap \KK^n).$$
\end{lemma}
\begin{proof}
    Using Definition \ref{DefJurriusPelik}, we get $M_k(\Cc) = \underset{\substack{\mathcal{\mathcal{D}} \subset \Cc \\ \dim (\mathcal{\mathcal{D}}) = k}}{\min} \text{wt}_R(\mathcal{\mathcal{D}}) = \text{wt}_R(\Cc)$. 

    By \cite[Remark 2.12]{BerhuyFaselGarotta}, we have ${\rm Rsupp}(\Cc) =\left(\Cc^{\perp}\cap\KK^n\right)^{\perp}$. Hence, we get $${\rm wt}_R(\Cc) = \dim_{\KK}(\text{Rsupp}(\Cc)) = \dim_{\KK}((\mathcal{C}^{\perp}\cap \KK^n)^{\perp}) = n- \dim_{\KK}(\Cc^{\perp}\cap \KK^n).$$
\end{proof}

In order to apply this result to compute the last rank distance of $\Cc$, we need to determine $\Cc^\perp$.

Recall now that if $M$ is a cyclic matrix, then $M^t$ is also a cyclic matrix (one way to check this quickly is to use the standard fact that an $n\times n$ matrix is cyclic if and only if its minimal polynomial has degree $n$). 

Therefore, the next proposition makes sense (see also \cite{lininv}).

\begin{proposition}\label{hdual}
Let $M\in{\rm M}_n(\KK)$ be a cyclic matrix with minimal polynomial $f\in\KK[x]$. If  $\Cc$ is an $M$-cyclic code, then $\Cc^\perp$ is an $M^t$-cyclic code. 

More precisely, if $f=gh$, where $g\in\LL[x]$ is the generator polynomial of $\Cc$, then $h$ is the generator polynomial of $\Cc^\perp$.  
\end{proposition}

\begin{proof}
We first have to show that if $\cv'\in \Cc^\perp$, then so is $\cv' (M^t)^t=\cv' M$.

Assume that $\cv'\in\LL^n$ satisfies $\cv'\cv^t=0$ for all $\cv\in \Cc$. 
Then, for all $\cv\in\Cc$, we have $$(\cv' M)\cv^t=\cv'(\cv M^t)^t=0,$$
since $\cv M^t\in\Cc$ by definition of an $M$-code. 
Hence, $\Cc^\perp$ is an $M^t$-cyclic code, as required. 

Now, let ${\bf v}, {\bf w}\in\KK^n$ be cyclic vectors for $M$ and $M^t$ respectively. Keeping the notation of the proposition, if $g$ has degree $n-k$, then $\Cc=\Cc_g$ has dimension $k$. Consequently, $\Cc^\perp$ has dimension $n-k$. Note now that $h$ has degree $k$, so that $\Cc_h$  also has dimension $n-k$. Hence, to prove that $h$ is the generator polynomial of $\Cc^\perp$, it is enough to prove that $\Cc_h\subset \Cc^\perp=\Cc_g^\perp$.

But, for all $Q_1,Q_2\in\LL[x]$, we have $${\bf w}h(M^t)^t Q_1(M^t)^t ({\bf v}g(M)^tQ_2(M)^t)^t={\bf w} (hQ_1Q_2g)(M) {\bf v}^t={\bf w}(Q_1Q_2f)(M){\bf v}^t=0,$$ since $f$ is the minimal polynomial of $M$.
This concludes the proof.
\end{proof}

We may now state and prove the main theorem of this subsection.

\begin{theorem}\label{ThmDernierDistance}
Let $M\in {\rm M}_n(\KK)$ be a cyclic matrix with minimal polynomial $f\in\KK[x]$. Write $f=f_1^{m_1}\cdots f_s^{m_s}$, where $f_1,\ldots,f_s$ are pairwise distinct monic irreducible polynomials of $\KK[x]$, and $m_1,\ldots,m_s\geq 1$.

    Let $\Cc\subset \LL^n$ be an $M$-cyclic code of dimension $k$, and let $g$ be its generator polynomial.

Write $g=g_1\cdots g_s$, where $g_i\mid f_i^{m_i}$ in $\LL[x]$.

For $i\in\left\llbracket 1,s\right\rrbracket$, set $$\ell_i=\left\lbrace\begin{array}{cl}
  0 &\mbox{ if }g_i=1 \cr \min\{\ell\in\left\llbracket 1,m_i\right\rrbracket\mid\ \  g_i\mid f_i^{\ell_i}\}& \mbox{ if }g_i\neq 1 \end{array}\right.,$$

  and $$\ell'_i= \left\lbrace\begin{array}{cl}0 &\mbox{ if }g_i=f_i^{m_i} \cr \min\{\ell'\in\left\llbracket 1,m_i\right\rrbracket\mid\ \  f_i^{m_i-\ell'}\mid g_i \}& \mbox{ if }g_i\neq f_i^{m_i}\end{array}\right. .$$

    Then we have  $$M_k(\Cc) =\sum_{i=1}^s \ell'_i \deg(f_i)=\sum_{\substack{1\leq i \leq s \\ g_i\neq f_i^{m_i}}}\ell'_i\deg(f_i),$$
    as well as 
    $$M_{n-k}(\Cc^{\perp}) =   \sum_{i=1}^s \ell_i \deg(f_i)= \sum_{\substack{1\leq i \leq s \\ {\rm gcd}(g, f_i) \neq 1}}\ell_i \deg(f_i).$$
\end{theorem}

\begin{proof}
The previous lemma shows that $M_{n-k}(\Cc^\perp) = n - \dim_{\KK}(\Cc\cap \KK^n)$. The second equality is then a direct application of Theorem \ref{Mcequal1}.

Now, recall from Proposition \ref{hdual} that $\Cc^\perp$ is an $M^t$-cyclic code, with generator polynomial $h$, where $f=gh$. Write $h=h_1\cdots h_s$, where $h_i\mid f_i^{m_i}$ in $\LL[x]$. Then we have $h_i=1$ if and only if $g_i=f_i^{m_i}$ and furthermore, for all $\ell'\in \left\llbracket 1, m_i \right\rrbracket$, we have $h_i\mid f_i^{\ell'}$ if and only if $f_i^{m_i-\ell'}\mid g_i$.

Since $M^t$ has minimal polynomial $f$, we may apply Theorem \ref{Mcequal1} to get the first equality.
\end{proof}

\begin{corollary}
Let $M\in {\rm M}_n(\KK)$ be a cyclic matrix with minimal polynomial $f\in\KK[x]$. Write $f=f_1^{m_1}\cdots f_s^{m_s}$, where $f_1,\ldots,f_s$ are pairwise distinct monic irreducible polynomials of $\KK[x]$, and $m_1,\ldots,m_s\geq 1$.

    Let $\Cc\subset \LL^n$ be an $M$-cyclic code of dimension $k$, and let $g$ be its generator polynomial.
    Then we have  $M_k(\Cc) = n$ if and only if $f_i\nmid g$ for all $i\in\left\llbracket 1,s\right\rrbracket$.
\end{corollary}

\begin{proof}
  With the notation of the previous theorem, we have $M_k(\Cc)=n$ if and only if $\ell'_i=m_i$ for all $i\in\left\llbracket 1,s\right\rrbracket$. Now, note that $\ell'_i<m_i$ if and only if $f_i\mid g_i$. The result follows, taking into account that $f_i\mid g_i$ is equivalent to $f_i\mid g$.  
\end{proof}

\begin{example}\label{exmkMcyclicctd}
Let us continue Example \ref{exmkMcyclic}. 

Recall that $\KK=\FF_3, \LL=\FF_{3^{10}}$,  $M=C_f\in {\rm M}_9(\KK)$, where $$f=(x^2+1)^2(x+1)^3(x-1)^2,$$ and that $g=(x-i)(x+1)^2(x-1)^2\in\LL[x]$, where  $i\in\LL$ satisfies $i^2=-1$. Then $\Cc_g$ has dimension $4$.

Since $f_1=x^2+1, f_2=x+1, f_3=x-1$, we have $\ell'_1=2, \ell'_2=1$ and $\ell'_3=0$, and thus $M_4(\Cc_g)=5$.

In particular, the bound proposed in Corollary \ref{coroweightMcyclic} is not sharp.
\end{example}
Once again, when $f$ is square-free, the results may be translated in a nicer way.

\begin{theorem}\label{ThmDernierDistancesqfree}
Let $M\in {\rm M}_n(\KK)$ be a cyclic matrix with minimal polynomial $f\in\KK[x]$. Assume that $f$ is square-free, and write $f=f_1\cdots f_s$, where $f_1,\ldots,f_s$ are pairwise distinct monic irreducible polynomials of $\KK[x]$.

    Let $\Cc\subset \LL^n$ be an $M$-cyclic code of dimension $k$, and let $g$ be its generator polynomial.

    Then we have  $$M_k(\Cc) = \sum_{\substack{1\leq i \leq s \\ f_i\nmid g}}  \deg(f_i)= n \ -  \sum_{\substack{1\leq i \leq s \\ f_i\mid g}}  \deg(f_i),$$
    as well as 
    $$M_{n-k}(\Cc^{\perp}) =  \sum_{\substack{1\leq i \leq s \\ {\rm gcd}(g, f_i) \neq 1}}\hspace{-0.3cm}\deg(f_i)= n \ - \hspace{-0.35cm} \sum_{\substack{1\leq i \leq s \\ {\rm gcd}(g, f_i) =1}}\hspace{-0.3cm}\deg(f_i).$$
\end{theorem}

\begin{example}
    Let $\KK=\FF_7$ and $\LL=\FF_{7^4}$, and let us consider the $[4,2]$-cyclic code $\Cc \subset \FF_{7^4}^4$ generated by $g = (x-1)(x-(4\omega^2-2))$, where $\omega$ is a generator of the cyclic group $\FF_{7^4}^\times$ satisfying $\omega^4 = \omega^2+1$. In particular, $(1,\omega,\omega^2,\omega^3)$ is an $\FF_7$-basis of $\FF_{7^4}$. 

We have $x^4-1=(x-1)(x+1)(x^2+1)\in\FF_7[x]$.

    Since ${\rm gcd}(g,x+1)=1$, we have $M_1(\Cc)=1$ by Theorem \ref{Mcequal1}. 
     Now, the only monic irreducible divisor of $x^4-1$ in $\FF_7[x]$ dividing $g$ is $x-1$, so Theorem \ref{ThmDernierDistancesqfree} gives us $M_2(\Cc) = 4 - \deg(x-1) = 3$. 
    
    One may recover this result directly as follows. Any codeword $\cv\in\Cc$ has the form $\cv=(a\alpha+b, -a(\alpha+1)+b\alpha, a-b\alpha,b)$ for some $a,b\in\LL$, where  $\alpha=4\omega^2-2$. Now, easy manipulations show that $${\rm Span}_\KK(a\alpha+b, -a(\alpha+1)+b\alpha, a-b\alpha,b)={\rm Span}_\KK(a\alpha, a-b\alpha,0,b).$$
    In particular, ${\rm wt}_R(\cv) \leq 3$ for all $\cv\in\Cc$. Moreover, for $a=\omega^2$ and $b=\omega$, one may check that $a\alpha, a-b\alpha$ and $b$ are $\KK$-linearly independent, so that the corresponding codeword has rank weight equal to $3$. By Remark \ref{remarkmaxrank}, we finally get that $M_2(\Cc)=3$.
\end{example}

\begin{remark}
When $f$ is square-free, the fact that $g_i\neq f_i$ is equivalent to say that $f_i\nmid g$, and the previous theorem shows that the bound of Corollary \ref{coroweightMcyclic} is sharp.    
\end{remark}

\section{Conclusion}

In this paper, we  studied the rank weight hierarchy for the so-called class of $M$-codes over an arbitrary field extension. The study of the generalized weights of this class is very relevant since it encompasses lots of well-known codes such as cyclic codes, quasi-cyclic codes and polynomial codes. 
We obtained upper bounds for the rank weight hierarchy of such codes, generalizing the work of \cite{Limoggier} for quasi-cyclic codes. Along the way, we derived a necessary condition for the existence of an MRD $M$-code in terms of the minimal polynomial of $M$, generalizing the fact that no cyclic codes are MRD. 
Finally, we studied a natural generalization of $f$-polynomial codes, namely $M$-cyclic codes, which corresponds to the case where $M$ is a cyclic matrix. We gave a necessary and sufficient condition for an $M$-cyclic code to have the first rank weight equals to $1$ in terms of its generator polynomial, and studied the proportion of such codes, with an application to cyclic and negacyclic codes. Finally, we obtained closed-form formulas for the last generalized rank weight of a $M$-cyclic code and its dual.

\section*{Acknowledgments}

The authors would like to express their sincere gratitude to F. Oggier for her valuable comments on an earlier version of this paper, and for suggesting us to investigate the last rank weight for $M$-cyclic codes. They also thank warmly the anonymous referees for their careful reading and insightful suggestions, which helped to improve the exposition of this work.




\begin{thebibliography}{100}
\bibitem{PolycyclicCode}
    A. Alahmadi, S. Dougherty, A. Leroy, P. Solé, {On the duality and the direction of polycyclic codes},
    \newblock {\it Advances in Mathematics of Communications,} \textit{10} (2016), 921--929.
    \url{https://www.aimsciences.org/article/id/fc8649f3-43a6-4ffd-90e4-91e18f8ed498}


\bibitem{Augot2013}
    D. Augot, P. Loidreau, G. Robert, {Rank metric and Gabidulin codes in characteristic zero}, 
    \newblock In {2013 IEEE International Symposium on Information Theory}, Istanbul, Turkey, (2013), 509--513
    \url{https://doi.org/10.1109/ISIT.2013.6620278}


\bibitem{Augot2017}
    D. Augot, P. Loidreau, G. Robert, {Generalized Gabidulin codes over fields of any characteristic}
    \newblock {\it Designs, Codes and Cryptography}, \textit{86}, (2017) 1807--1848
    \url{https://api.semanticscholar.org/CorpusID:1033838}


\bibitem{FnT}
   H. Bartz, L. Holzbaur, H. Liu, S. Puchinger, J. Renner, A. Wachter-Zeh, {Rank-Metric Codes and Their Applications},
   \newblock {\it Foundations and Trends® in Communications and Information Theory,}
   \newblock \textit{19} (2022), 390--546.
   \url{http://dx.doi.org/10.1561/0100000119}


\bibitem{BerhuyFaselGarotta}
    G. Berhuy, J. Fasel, O. Garotta, {Rank weights for arbitrary finite field extensions},
    \newblock {\it Advances in Mathematics of Communications,}
    \newblock \textit{15} (2019), 575--587.
    \url{https://api.semanticscholar.org/CorpusID:59600064}


\bibitem{JeromeDucoat}
    J. Ducoat, {Generalized rank weights: A duality statement},
    \newblock In {Topics in Finite Fields}, American Mathematical Society, G. Kyureghyan, G. L. Mullen and A. Pott Eds.
    \newblock \textit{632} (2015) 101--109.
    \url{https://api.semanticscholar.org/CorpusID:18454082}


\bibitem{DucOggier1}
    J. Ducoat, F. E. Oggier, {Rank weight hierarchy of some classes of polynomial codes},
    \newblock {\it Des. Codes Cryptography}
    \newblock \textit{91} (2022), 1627--1644.
    \url{https://doi.org/10.1007/s10623-022-01181-6}


\bibitem{Gabidulin}
    E. Gabidulin, {Theory of codes with maximum rank distance (translation)}, 
    \newblock {\it Problems of Information Transmission,}
    \newblock \textit{21} (1985), 1--12.

    
\bibitem{GhorpadeJohnsen}
    S. R. Ghorpade, T. Johnsen, {A polymatroid approach to generalized weights of rank metric codes},
    \newblock {\it Des. Codes Cryptography,}
    \newblock \textit{88} (2020), 2531--2546.
    \url{https://doi.org/10.1007/s10623-020-00798-9}


\bibitem{Graner}
    A.-M. Graner, {Closed formulas for the factorization of $x^n-1$, the $n$-th cyclotomic polynomial, $x^n-a$ and $f(x^n)$ over a finite field for arbitrary positive integers $n$},
    \newblock arxiv (2024).
    \newblock \url{https://arxiv.org/abs/2306.11183}

\bibitem{Huffman_Pless_2003}
    \newblock W. C. Huffman, V. Pless, {\em Fundamentals of Error-Correcting Codes},
    \newblock Cambridge University Press, 2003.
    \url{https://doi.org/10.1017/CBO9780511807077}


 \bibitem{JurriusPellikaan}
      R. Jurrius, R. Pellikaan, {On defining generalized rank weights},
     \newblock  {\it Advances in Mathematics of Communications,}
     \newblock  \textit{11} (2017), 225--235. \url{http://dx.doi.org/10.3934/amc.2017014}


\bibitem{KuriMatsuUye}
    J. Kurihara, R. Matsumoto, T. Uyematsu, {Relative Generalized Rank Weight of Linear Codes and Its Applications to Network Coding},
    \newblock {\it IEEE Transactions on Information Theory,}
    \newblock \textit{61} (2013), 3912--3936.
    \url{https://api.semanticscholar.org/CorpusID:2443740}


\bibitem{LidlNiederreiter}
    R. Lidl, H. Niederreiter, P. M. Cohn, {\em Finite fields},
    \newblock Second edition, Cambridge, Cambridge University Press, 1997.
    \url{https://doi.org/10.1017/CBO9780511525926}


\bibitem{Limoggier}
    E. Lim, F. E. Oggier, {On the generalised rank weights of quasi-cyclic codes},
    \newblock {\it Advances in Mathematics of Communications,}
    \newblock \textit{18(1)} (2024), 192--205.
    \url{https://doi.org/10.3934/amc.2022010}


\bibitem{MartinezP}
    U. Martínez-Pe{\~n}as, {Generalized Rank Weights of Reducible Codes, Optimal Cases, and Related Properties}, 
    \newblock {\it IEEE Transactions on Information Theory,}
    \newblock \textit{64} (2018), 192--204.


\bibitem{OggierSboui}
    F. E. Oggier, A. Sboui, {On the existence of generalized rank weights},
    \newblock {\it 2012 International Symposium on Information Theory and its Applications,}
    \newblock (2012), 406--410.  \url{https://api.semanticscholar.org/CorpusID:15579437}


\bibitem{lininv}
H. Ou-azzou, M.Najmeddine, N. Aydin, P. Liu,E. Ialou, M. El Mahdi, 
{Linear codes invariant under a linear endomorphism}
\newblock {\bf 19}, vol.2 (2025), 676--607.
\url{https://www.aimsciences.org/article/id/664005f1475da12c51d5e2b9}


\bibitem{roman}
S. Roman, {\em Advanced Linear Algebra. Third edition},
\newblock{Graduate texts in Mathematics {\bf 135}, Springer, 2008.}
\url{https://link.springer.com/book/10.1007/978-0-387-72831-5}


\bibitem{ROTH}
    R. M. Roth, {Maximum-rank array codes and their application to crisscross error correction},
    \newblock {\it IEEE Transactions on Information Theory,}
    \newblock \textit{37} (1991), 328--336.
    \url{http://dx.doi.org/10.1109/18.75248}


\bibitem{shiromoto}
    K. Shiromoto, {Codes with the rank metric and matroids},
    \newblock {\it Des. Codes Cryptography,}
    \newblock \textit{87} (2019), 1765--1776.
    \url{https://doi.org/10.1007/s10623-018-0576-0}



\end{thebibliography}
\end{document}